\documentclass{appolb}
\usepackage{epsf,rotate}
\usepackage{amsmath,amstext,amsbsy,amsopn,amsfonts,amssymb}
\usepackage[dvips]{graphicx}
\usepackage[latin1]{inputenc}
\usepackage[english]{babel}
\usepackage[T1]{fontenc}
\usepackage{psfrag}

\def \SL  {\mbox{SL}}

\def \Im {\mathop{\rm Im}\nolimits}
\def \Re {\mathop{\rm Re}\nolimits}

\newcommand{\Log}{\mbox{Log}}
\newcommand{\eqdef}{\mbox{$ \stackrel{\rm def}{=}$}}

\newcommand\lr[1]{{\left({#1}\right)}}

\newcommand \wbar [1] {\overline{#1}}

\newcommand \vev [1] {\langle{#1}\rangle}
\newcommand \ket [1] {|{#1}\rangle}

\newcommand \mybf[1] {\mbox{\boldmath$ {#1} $}}
\newcommand \ot[1] {\Hat{t}_{#1}}
\newcommand \otb[1] {\Hat{\wbar{t}}_{#1}}
\newcommand \oq[1] {\Hat{q}_{#1}}
\newcommand \oqb[1] {\Hat{\wbar{q}}_{#1}}

\def \bell {{\mybf \ell}}
\def \bellp {{\mybf \ell'}}

\begin{document}

\title{Four-particle solutions 
to Baxter equation of
$\SL(2,\mathbb{C})$ Heisenberg spin magnet for 
integer conformal Lorentz spin 
and their normalizability}
\author{Jan~Kotanski\footnote{Institute of Physics,
Jagellonian University, Reymonta 4, PL-30-059 Cracow,Poland}
}


\maketitle
                                                                  
\begin{abstract}
{\normalsize
The four reggeized gluon states for non-vanishing Lorentz conformal spin 
$n_h$ are considered. 
To calculate their spectrum the $Q-$Bax\-ter method is used.
As a result we describe normalizable trajectory-like states,
which form continuous spectrum,
as well as discrete point-like solutions,
which turn out to be non-normalizable. 
The point-like solutions exist
due to symmetry of the Casimir operator where conformal weights
 $(h,\wbar h)\rightarrow (h,1-\wbar h)$.
}
\end{abstract}

\newline

\vspace*{1cm}
\noindent TPJU-2/2007

%

\bibliographystyle{unsrt}

\section{Introduction}
\label{sec:intr}

In high energy limit of QCD the scattering processes can be described
by means of effective particles that are compound states of reggeized gluons,
shortly Reggeons 
\cite{gell,Gribov:1968fc,Fadin:1975cb}. 
In calculations of the scattering processes of hadrons
with $N-$Reggeon exchange one can find that the high energy asymptotics of 
the scattering amplitude is governed by intercepts \cite{Gribov:1968fc,
Fadin:1975cb}.
On the other hand, applying reggeized gluons to deep
inelastic processes one can compute anomalous dimensions of QCD
for the hadron structure function 
\cite{Jaroszewicz:1982gr,Korchemsky:2003rc,deVega:2002im}. 

To find the intercepts as well as the anomalous dimensions the
BKP equation is considered
\cite{Bartels:1980pe}, 
which is a generalization of the BFKL equation 
\cite{Fadin:1975cb}
to multi-Reggeon exchange. This equation 
has a structure of a Schr\"odinger equation and it corresponds to
the non-compact Heisenberg $\SL(2,\mathbb{C})-$spin magnet.
The energy in this equation is called the Reggeon energy
while the eigenfunctions are the Reggeon wave-functions.  
To solve the BKP equation one can equivalently 
calculate solutions of
the Baxter $Q-$operator 
eigenproblem \cite{Lipatov:1994xy,Faddeev:1994zg} defined by Baxter equations
\cite{Derkachov:2001yn,Derkachov:2002wz,Derkachov:2002pb}. 
The system is completely integrable, it possesses a complete set of integrals
of motion, \ie conformal charges $\{q_k, \wbar q_k\}$ with $k=2,\ldots,N$.
Therefore, the energy is a function of these conformal charges.
Eigenvalues of the lowest conformal charges, \ie $(q_2,\wbar q_2)$,
depend on conformal weights $(h,\wbar h)$ which
are parameterized by
the real scaling dimension $\nu_h$ and the integer conformal
Lorentz spin $n_h$. Additionally,
applying the WKB approach  \cite{Derkachov:2002pb},
which consists in constructing the asymptotic solutions to Baxter equations,  
one finds that the remaining conformal charges
are parameterized by additional set of $(2N-4)-$integer parameters.
For fixed $(h,\bar h)$ these integer numbers enumerate   
vertices of WKB lattices 
which correspond to higher conformal charges 
\cite{Derkachov:2002pb,Kotanski:2006sh}.
Since the solutions depend on continuous parameter $\nu_h$ 
the remaining conformal charges $q_3,\ldots,q_N$ form
one-dimensional trajectories in the conformal charge space.

In this work
we focus on four-Reggeon exchanges which have been described in
the following references \cite{Derkachov:2002wz,Kotanski:2006sh,Kotanski:2005ci,
deVega:2002im}. The intercept of these exchanges
is defined by the minimal energy as
$\alpha(0)=1- \min(E_4) \alpha_s N_c/(4 \pi)$.
Contrary to Ref.~\cite{Derkachov:2002wz} 
the authors 
of Ref.~\cite{deVega:2002im-bis} claim
that the leading contribution in the $N=4$ Reggeon  
sector comes from the exchange 
with $n_h=2$ with the intercept  higher than 
the intercept of the BFKL-pomeron.
Here we intend to explain these results. 
Thus, we analyse the energy spectrum of the exchanged states and 
calculate the spectrum of the conformal charges with $n_h\in \mathbb{Z}$. 
Indeed, it turns out that additionally to the ordinary trajectory states, 
there exist other solutions to the Baxter equations.
These solutions appear only for $n_h \ne 0$
and are not situated on any trajectory, \ie they are point-like.
Moreover, their conformal charges correspond to the results of 
Refs.~\cite{deVega:2002im,deVega:2002im-bis}.

In order to contribute to the scattering amplitude processes, 
\ie to be physical, the solutions should be normalizable according to the 
$\SL(2,\mathbb{C})$ scalar product. 
Thus, the scalar product of 
the trajectory solutions with continuous parameters $\nu_h$ and $\nu_h'$
has to be proportional to $\delta(\nu_h-\nu_h') \delta_{n_h n_h'}$,
while the scalar products 
of the point-like solutions to the Kronecker delta $\delta_{n_h n_h'}$.
On the other hand, by construction, 
the wave functions of the eigenstates should have
transformation property with respect to $\SL(2,\mathbb{C})$ 
conformal transformations, 
\ie they have to
be homogeneous functions of two-dimensional coordinates and must not involve 
any scale. Therefore, the product of two eigenstates could be either 0 
or $\infty$.
If the eigenstates are continuous in $\nu_h$
(trajectory-like solutions),
then 0 or $\infty$ follows form the Dirac delta function.
If the eigenstates are discrete either 
the norm of each eigenstate is infinite
and this means that this state does not contribute to the decomposition of the 
Hamiltonian over the eigenstates, or
the eigenstates and/or the scalar product involve some scale, which
breaks conformal symmetry.
However, as will be shown in the following, 
the point-like solutions have infinite norm so these
solutions 
are not normalizable w.r.t.~$\SL(2,\mathbb{C})$ scalar product.

The existence of non-normalizable states may be understood by
performing analytical continuation of the physical trajectory solutions 
from the $n_h=0$ sector to the complex $\nu_h-$plane
which implies relaxation of the normalization condition.
It turns out that due to a symmetry of the Casimir operator,
solutions from this analytical continuation plane with 
$i \nu_h \in \mathbb{Z}/2$ have their twin-solutions
with $\nu_h=0$ and $n_h \in \mathbb{Z}$.
These twin-solutions have the same conformal charges
as the corresponding solutions from the $n_h=0$ 
analytical plane.
They differ only in the conformal weight with $\wbar h \to 1-\wbar h$. 
The latter transformation may move states outside the normalizable
physical range.
Therefore, some twin-solutions, \ie point-like ones,
as will be shown in the following, are non-normalizable.

%

The energy of the ground state gives the largest contribution
to the scattering amplitude. Assuming that this ground state is not 
degenerated
implies $q_3=0$
\cite{Korchemsky:1994um}.
Usually the trajectory solutions have the lowest energy at $\nu_h=0$. 
Moreover, the energy of the trajectory solution with $n_h\ne0$ is always  
higher than the minimal energy of the states from the $n_h=0$ sector.
For even $n_h$ and $q_3=0$
the normalizable solutions coexist with 
non-normalizable ones. Here, the point-like solutions do not correspond 
to WKB lattices and their number for fixed $n_h$ equals $|n_h/2|$.
The point-like  solution with the lowest energy 
belongs to the $n_h=4$ sector and 
its energy is comparable to two-Pomeron exchange energy.
On the other hand 
we have found that all solutions with odd $n_h$ and $q_3=0$ 
are point-like and non-normalizable,
and can be described by vertices of WKB lattices.

In this work we compute numerically the Hamiltonian
spectra of four reggeized gluons.
As a result we get trajectory solutions 
to the Baxter equations which form
continuous spectra.
Moreover, we also find the point-like solutions.
We try to look more closely at their origin and discuss
their normalizability conditions.
In Section \ref{sec:Ham} the Reggeon Hamiltonian is introduced and the
way of finding its spectrum using the Baxter method is shown. 
In the next section the integral ansatz 
is applied to rewrite the Baxter equations
in a  differential form and find their solutions.
Moreover, the $\SL(2,\mathbb{C})-$norm is evaluated.
In Section \ref{sec:traj} 
the normalizable trajectory-like solution are presented
and analytical continuation to the complex $\nu_h-$plane is performed. 
Furthermore, the symmetry between solutions with
(half-)integer conformal charges is established.
In Section \ref{sec:tsol} 
trajectory-like solutions at the specific point $\nu_h=0$ 
are discussed. 
Furthermore, in Section \ref{sec:psol} 
non-normalizable point-like solutions are constructed.
They also appear at $\nu_h=0$. However, they are not
situated on any trajectory and 
we check they are non-normalizable. 
At the end we add final conclusions.

\section{Hamiltonian and the Baxter operator}
\label{sec:Ham}
\subsection{Hamiltonian}

In the  high energy Regge limit the  total energy $s \to \infty$ and 
the four-momentum transfer $t$ is constant.
In this limit  the contribution of $N=4$
reggeized gluons to the scattering amplitude 
can be written as:
\begin{equation}
\mathcal{A}_{N=4}(s,t) = s \int d^2 z_0\,
\e^{i\vec z_0 \cdot \vec p}
\vev{{\Phi}_A(\vec z_0)|
\e^{- \wbar{\alpha}_s Y {\cal H}_4/4 }
 | {\Phi}_B(0)}\,,
\label{A}
\end{equation}
where the rapidity $Y=\ln s$ and
$\wbar{\alpha}_s=\alpha_s N_c/\pi$ is the strong coupling constant.
The wave-functions 
$\ket{\Phi_{A(B)}(\vec z_0)}\equiv \Phi_{A(B)}(\vec z_i-\vec z_0)$
describe the coupling of four gluons to the scattered particles.
The $\vec z_0 -$ integration fixes the momentum transfer $t=-\vec p^{\,2}$.
In the multi-colour limit \cite{'tHooft:1973jz} 
the Hamiltonian   
\begin{equation}
\mathcal{H}_{4}=H_{4}+\wbar{H}_{4}\mbox {,}\qquad 
[H_{4},\wbar{H}_{4}]=0\,.
\label{eq:sepH}
\end{equation}
Holomorphic and anti-holomorphic Hamiltonians can be written
in the M\"obius representation 
\cite{Lipatov:1993yb,Faddeev:1994zg}
as 
\begin{equation}
H_{4}=\sum _{k=1}^{4}H(J_{k,k+1})\mbox {,}\quad 
\wbar{H}_{4}=\sum _{k=1}^{4}H(\wbar{J}_{k,k+1})\mbox {,}
\label{eq:Ham}
\end{equation}
where
\begin{equation}
H(J)=\psi (1-J)+\psi (J)-2\psi (1)\,,
\end{equation}
with $\psi (x)=d\ln \Gamma (x)/dx$ being the Euler digamma function
and $J_{4,5}=J_{4,1}$. The operators, $J_{k,k+1}$ and $\wbar{J}_{k,k+1}$,
are defined through the Casimir operators for the sum of the 
$\SL(2,\mathbb{C})$ spins related to
the neighbouring Reggeons \cite{Derkachov:2001yn}
\begin{equation}
J_{k,k+1}(J_{k,k+1}-1)=(S^{(k)}+S^{(k+1)})^{2}\,,
\label{eq:jvss}
\end{equation}
with $S_{\alpha }^{(5)}=S_{\alpha }^{(1)}$, and $\wbar{J}_{k,k+1}$ is
defined similarly. 

The high energy behaviour of the scattering amplitude  (\ref{A})
is governed by 
the eigenvalues of ${\cal H}_4$,  which are also called Reggeon energies,
$E_{4}(q,\wbar q)$.
In order to find $E_{4}(q,\wbar q)$ 
one has to solve the Schr\"{o}dinger equation 
\begin{equation}
\mathcal{H}_{4}
\Psi_{\vec{p},\{q,\wbar q\}} (\vec{z}_{1},\vec{z}_{2},\vec{z}_{3} ,\vec{z}_{4})
=E_{4}(q,\wbar q)\Psi_{\vec{p},\{q,\wbar q\}} (\vec{z}_{1},\vec{z}_{2},\vec{z}_{3} ,\vec{z}_{4})\,,
\label{eq:Schr}
\end{equation}
where different energy values are enumerated by quantum numbers 
$\{q,\wbar q\}\equiv(q_2,q_3,q_4,\wbar q_2, \wbar q_3,\wbar q_4)$.
The eigenstates $\Psi_{\vec{p},\{q,\wbar q\}} 
(\vec{z}_{1},\vec{z}_{2},\vec{z}_{3} ,\vec{z}_{4})$
are single-valued functions on the planes $\vec{z_i}=(z_i,\wbar{z_i})$,
normalizable with respect to the $\SL(2,\mathbb{C})$ invariant scalar
product\begin{equation}
\vev{\Psi_{\vec{p},\{q,\wbar q\}}|\Psi_{\vec{p},\{q,\wbar q\}}}=
\int d^{2}z_{1}d^{2}z_{2}d^{2}z_{3}d^{2}z_{4}
|\Psi_{\vec{p},\{q,\wbar q\}} (\vec{z}_{1},\vec{z}_{2} ,\vec{z}_{3} ,\vec{z}_{4})|^{2}\,,
\label{eq:norm}
\end{equation}
where $d^{2}z_i=dx_idy_i=dz_id\wbar{z}_i/2$ with $\wbar z_i={z_i}^{\ast}$\footnote{Here {\em bar} is used to denote quantities in the anti-holomorphic sector, whereas the {\em asterisk} denotes complex conjugation}.

\subsection{Baxter operator}

In order to solve the Schr\"odinger equation (\ref{eq:Schr}) 
one can apply
the powerful method 
of the Baxter $Q$-operator \cite{Baxter,Derkachov:2001yn}.
This operator depends 
on two complex spectral parameters $u$, $\wbar u$ and  
in the following will be denoted as
$\mathbb{Q}(u,\wbar{u})$.  
By definition $\mathbb{Q}(u,\wbar{u})$
has to satisfy the commutativity relations
\begin{equation}
\left[ \mathbb{Q} (u,\wbar{u}) ,
 \mathbb{Q}(v,\wbar{v}) \right]=0\,,
\label{eq:comQQ}
\end{equation}
and
\begin{equation}
\left[\ot{4}(u), \mathbb{Q}(u,\wbar{u}) \right]=
\left[\otb{4}(\wbar{u}), \mathbb{Q}(u,\wbar{u}) \right]=0\,,
\label{eq:comQt}
\end{equation}
as well as Baxter equations
\begin{equation}
\ot{4}(u) \mathbb{Q}(u,\wbar{u})  =
(u + i s)^4 \mathbb{Q}(u+i,\wbar{u})  +
(u - i s)^4 \mathbb{Q}(u-i,\wbar{u})  \,,
\label{eq:Baxeq}
\end{equation}
\begin{equation}
\otb{4}(\wbar{u}) \mathbb{Q}(u,\wbar{u})  =
(\wbar{u} + i \wbar{s})^4 \mathbb{Q}(u,\wbar{u}+i)  +
(\wbar{u} - i \wbar{s})^4 \mathbb{Q}(u,\wbar{u}-i)  \,,
\label{eq:Baxbeq}
\end{equation}
where  the auxiliary transfer matrices 
\begin{equation}
\ot{4}(u)
=2u^{4}+\oq{2} u^{2}+ \oq{3} u +\oq{4}\,,
\label{eq:tnu}
\end{equation}
and similarly for $\otb{4}(\wbar{u})$, are polynomials 
in spectral parameter $u$ with the coefficients being the 
integrals of motion, \ie conformal charges. 
In the QCD case the complex spins $(s,\wbar s)=(0,1)$.

The operators of conformal charges have 
a particularly simple form for the $\SL(2,\mathbb{C})$ spins $s=0$:
\begin{equation}
\oq{k}=i^k\sum_{1\le j_1 < j_2 < \ldots  < j_k\le 4}
z_{j_1j_2}\ldots z_{j_{k-1},j_k}z_{j_k,j_1}\partial_{z_{j_1}}\ldots 
\partial_{z_{j_{k-1}}}\partial_{z_{j_k}}\,,
\label{eq:qks0}
\end{equation}
as well as for $\wbar s=1$:
\begin{equation}
\oqb{k}=i^k\sum_{1\le j_1 < j_2 < \ldots  < j_k\le 4}
\partial_{\wbar z_{j_1}}\ldots \partial_{\wbar z_{j_{k-1}}}\partial_{\wbar z_{j_k}}
\wbar z_{j_1j_2}\ldots \wbar z_{j_{k-1},j_k} \wbar z_{j_k,j_1}\,.
\label{eq:qks1}
\end{equation}
with $z_{ij}=z_i-z_j$ and $\wbar z_{ij}=\wbar z_i-\wbar z_j$ .
Eigenvalues of the lowest conformal charges, 
\begin{equation}
q_2=-h(h-1),\quad
\mbox{and}\quad  \wbar q_2=-\wbar h (\wbar h -1),
\label{eq:evq2}
\end{equation}
are parameterized
by conformal weight 
\begin{equation}
h=\frac{1+n_{h}}{2}+i\nu _{h}\mbox {,}\qquad \wbar h=
\frac{1-n_{h}}{2}+i\nu _{h}\,,
\label{eq:hpar}
\end{equation}
where integer $n_h$ has a meaning of the two-dimensional
Lorentz spin, 
whereas $\nu_{h}$ defines the
scaling dimension. 
Note that the complex spins $(s,\wbar{s})$ and 
conformal weights $(h,\wbar{h})$ parameterize
the irreducible representations of the $\SL(2,\mathbb{C})$ group.

The Baxter $\mathbb{Q}(u,\wbar{u})$-operator
as well as the Hamiltonian 
(\ref{eq:sepH}) commute with the auxiliary transfer matrices (\ref{eq:comQt}),
and as a consequence
they share  
the common set of eigenfunctions.
Moreover, the eigenvalues of the $Q$-operator, defined by
\begin{equation}
 \mathbb{Q}(u,\wbar{u}) 
\Psi _{\vec{p},\{q,\wbar q\}}(\vec{z}_{1},\vec{z}_{2},\vec{z}_{3},\vec{z}_{4}) 
=  Q_{q,\wbar q}(u,\wbar{u}) 
\Psi _{\vec{p},\{q,\wbar q\}}(\vec{z}_{1},\vec{z}_{2},\vec{z}_{3},\vec{z}_{4})\,,
\label{eq:eeqQ}
\end{equation} 
 satisfy the same Baxter equations
(\ref{eq:Baxeq}) and (\ref{eq:Baxbeq})
with the auxiliary transfer matrices replaced by their corresponding 
eigenvalues.
Thus, to find the eigenvalues of the $Q-$operator 
we solve the eigenproblem 
of the Hamiltonian (\ref{eq:sepH}), which
can be written in terms of 
the Baxter $Q$-operator \cite{Derkachov:2001yn}:
\begin{equation}
{\cal H}_4=\epsilon_4 + \left. i \frac{d}{du} 
\ln \mathbb{Q}(u+is,\wbar{u}+i \wbar{s})\right|_{u=0}
-\left( \left. i \frac{d}{du} 
\ln \mathbb{Q}(u-is,\wbar{u}-i \wbar{s})\right|_{u=0}\right)^{\dagger}\,,
\label{eq:HNQQ}
\end{equation}
where the additive normalization constant is given as
\begin{equation}
\epsilon_4 = 8 \, \Re [\psi(2 s)+ \psi(2-2s)-2 \psi(1)]\,.
\label{eq:epsnor}
\end{equation}

Finally, the Hamiltonian 
(\ref{eq:sepH}) 
is invariant
under cyclic and mirror permutations \cite{Derkachov:2001yn}
defined by
\begin{eqnarray}
\nonumber
 \mathbb{P}\Psi _{\{q,\wbar q\}}(\vec{z}_{1},\vec{z}_{2},\ldots,\vec{z}_{4}) &
\eqdef & \Psi _{\{q,\wbar q\}}(\vec{z}_{2},\vec{z}_{3},\ldots,\vec{z}_{1})\\
\nonumber
&& = e^{i\theta_4 (q,\wbar q)}\Psi _{\{q,\wbar q\}}(\vec{z}_{1},\vec{z}_{2},
\dots,\vec{z}_{4})\,,\\
\nonumber
 \mathbb{M}\Psi ^{\pm }(\vec{z}_{1},\vec{z}_{2},\ldots,\vec{z}_{4}) &
\eqdef & \Psi ^{\pm }(\vec{z}_{4},\vec{z}_{3},\ldots,\vec{z}_{1}) \\
&& =  \pm \Psi ^{\pm }(\vec{z}_{1},\vec{z}_{2},\ldots,\vec{z}_{4})\,.
\label{eq:PMsym}
\end{eqnarray}
where $\pm q=(q_2,\pm q_3, q_4)$ and
\begin{equation}
\Psi ^{\pm }(\vec{z}_{1},\vec{z}_{2},\ldots,\vec{z}_{4})=
\frac{1}{2}\left(\Psi_{\{q,\wbar q\}}(\vec{z}_{1},\vec{z}_{2},\ldots,\vec{z}_{4}) 
\pm \Psi_{\{-q,-\wbar q\}}(\vec{z}_{1},\vec{z}_{2},\ldots,\vec{z}_{4})\right)\,.
\label{eq:MPsi}
\end{equation}
The eigenvalues of $\mathbb{P}$ depend on
quasimomentum $\theta_N=0,\frac{\pi}{2}, \pi, \frac{3 \pi}{2}$.
According to the Bose symmetry of the reggeized gluons $\theta_4=0$ 
\cite{Derkachov:2002wz}.

\subsection{Normalization of the Baxter solution}

It turns out that the norm (\ref{eq:norm}) can be also rewritten in terms of 
eigenvalues of $\mathbb{Q}(u,\wbar u)$ \cite{Derkachov:2001yn}.
The $\SL(2,\mathbb{C})$ scalar product is normalized as
\begin{multline}
 \int
\Psi_{\vec{p},\{q,\wbar q\}}
(\vec{z}_1,\vec{z}_2,\vec{z}_3,\vec{z}_4)
\left(\Psi_{\vec{p}',\{q',\wbar q'\}}(
\vec{z}_1,\vec{z}_2,\vec{z}_3,\vec{z}_4)\right)^{\ast}
\prod_{k=1}^{4}d^{2}z_{k} =\\
=(2\pi)^{4}\delta^{(2)}(\vec{p}-\vec{p}^{\, '})
\delta_{\{q,\wbar q\}\{q,\wbar q\}'}\,,
\label{eq:nrmq}
\end{multline}
where $q=(q_{2},q_{3},q_{4})$  and anti-holomorphic $\wbar q=(\wbar q_{2},\wbar q_{3},\wbar q_{4})$ depend on
real $\nu_{h}$, integer $n_{h}$ and on the set of integers 
$\bell=(\ell_{1},\ell_{2},\ell_3,\ell_4)$.
The Reggeon wave-function in Sklyanin representation (SoV) \cite{Sklyanin:1991ss,Derkachov:2001yn} 
can be written 
using separated coordinates
$\vec{\mybf x}=(\vec{x}_1,\vec{x}_2,\vec{x}_{3})$ as
\begin{multline}
\Psi_{\vec{p},\{q,\wbar q\}}(\vec{z}_1,\vec{z}_2,\vec{z}_3,\vec{z}_4)=\\
=\int d^{3}\vec{\mybf x}\mu(\vec{x}_{1},\vec{x}_2,\vec{x}_{3})U_{\vec{p},\vec{\mybf x}}(\vec{z}_{1},\vec{z}_2,\vec{z}_3,\vec{z}_{4})(\Phi_{\{q,\wbar q\}}(\vec{x}_{1},\vec{x}_{2},\vec{x}_{3}))^{\ast}\:,
\label{eq:phi}
\end{multline}
where $\vec{\mybf x}=(\vec{x}_{1},\vec{x}_{2},\vec{x}_{3})$ are separated
variables, $U_{\vec{p},\vec{\mybf x}}(\vec{z}_{1},\ldots,\vec{z}_{4})$
is unitary transformation while
\begin{equation}
\left(\Phi_{\{q,\wbar q\}}(\vec{x}_{1},\vec{x}_2,\vec{x}_{3})\right)^{\ast}
=e^{i\theta_{4}(q,\wbar q)/2}\prod_{k=1}^{3}
\left(
\frac{\wbar x_k}{x_k}
\right)^{4}
Q_{q,\wbar q}(x_{k},\wbar{x}_{k})\,,
\end{equation}
and $Q_{q,\wbar q}(x_{k},\wbar{x}_{k})$ is the eigenvalue of the Baxter operator.
The separated coordinates 
\begin{equation}
x_{k}=\nu_{k}-\frac{in_{k}}{2}\,,\qquad\wbar{x}_{k}=\nu_{k}+\frac{in_{k}}{2}\,,
\end{equation}
are parameterized by $n_{k}$ integer and $\nu_{k}$ real\footnote{The parameters
$n_k$ and $\nu_k$ should not be identified with $n_h$ and $\nu_h$ from 
Eq.~(\ref{eq:hpar}).}. 
Integration over their space implies
\begin{equation}
\int d^{3}\vec{\mybf x}=\prod_{k=1}^{3}
\sum_{n_{k}=-\infty}^{\infty}\int_{-\infty}^{\infty}d\nu_{k},
\qquad\mu(\vec{\mybf x})=\frac{2}{6 \pi^{16}}\prod_{j,k=1,j>k}^3
\left|\vec{x}_{k}-\vec{x}_{j}\right|^{2}\,,
\end{equation}
where $\left|\vec{x}_{k}-\vec{x}_{j}\right|^{2}=(\nu_{k}-\nu_{j})^{2}+(n_{h}-n_{j})^{2}/4$.
The unitary integral kernel 
$U_{\vec{p},\vec{\mybf x}}$ can be rewritten
\begin{equation}
U_{\vec{p},\vec{\mybf x}}(\vec{z}_{1},\ldots,\vec{z}_{4})
=c_{4}(\vec{\mybf x})(\vec{p}^{\,\,2})^{3/2}\int d^{2}z_{0}\,
e^{2i\vec{p}\cdot\vec{z}_{0}}U_{\vec{\mybf x}}(\vec{z}_{1},\ldots,
\vec{z}_{4};\vec{z}_{0})\,,
\label{eq:Upker}
\end{equation}
with $2\vec{p}\cdot\vec{z}_{0}=p\, z_{0}+\wbar{p}\,\wbar{z}_{0}$ and
it is normalized to
\begin{multline}
\int d^{4}\vec{z}\, U_{\vec{p},\vec{\mybf x}}(\vec{z}_{1},\ldots,\vec{z}_{4})
\left(U_{\vec{p}^{\, '},\vec{\mybf x}'}
(\vec{z}_{1},\ldots,\vec{z}_{4})\right)^{\ast}=\\
=(2\pi)^{4}\delta^{(2)}(\vec{p}-\vec{p}^{\, '})
\{\delta(\vec{\mybf x}-\vec{\mybf x}')+\ldots\}
\frac{1}{6 \mu(\vec{\mybf x})}\:,
\end{multline}
where $\vec{z}=(\vec{z}_{1},\ldots,\vec{z}_{4})$ and ellipses denote
the sum of terms involving all permutations of the vector inside the
$\vec{\mybf x}=(\vec{x}_{1},\vec{x}_{2},\vec{x}_{3})$. 
The kernel  $U_{\vec{\mybf x}}$ is defined in 
Ref.~\cite{Derkachov:2001yn}.
Substituting the above formulae, the scalar product 
of the eigenstates (\ref{eq:nrmq})
is given by
\begin{multline}
(2\pi)^{4}\delta^{(2)}(\vec{p}-\vec{p}^{\, '})e^{i(\theta_{4}(q,\wbar q)-\theta_{4}(q',\wbar q'))/2}\int\prod_{k=1}^{3}d^{2}x_{k}^{\ast}\left(\mu(\vec{x}_{1},\vec{x}_{2},\vec{x}_{3})\right)^{\ast} \\
\times Q_{q,\wbar q}(x_{k},\wbar{x}_{k})(Q_{q',\wbar q'}(x_{k},\wbar{x}_{k}))^{\ast}
 =(2\pi)^{4}\delta^{(2)}(\vec{p}-\vec{p}^{\, '})\delta_{\{q,\wbar q\}\{q',\wbar q'\}}\,.
\label{eq:normQQ}
\end{multline}
where $d^{2}x_{k}^{\ast}\left(\mu(\vec{x}_{1},\vec{x}_{2},\vec{x}_{3})\right)^{\ast}=d^{2}x_{k}\mu(\vec{x}_{1},\vec{x}_{2},\vec{x}_{3})$
and 
$\delta_{\{q,\wbar q\}\{q,\wbar q\}'}
=\delta(\nu_{h}-\nu_{h}')\delta_{n_{h}n_{h}'}\delta_{\bell\bellp}$. 
The quasimomentum $\theta_4$ depends only on integer parameters, \ie $n_{h},\bell$,
therefore for the orthonormal states the normalization condition
can be written up to the momentum term as
\begin{multline}
\langle q,\wbar q|q',\wbar q'\rangle
\eqdef (2\pi)^{4}\int\prod_{k=1}^{3}d^{2}x_{k}\mu(\{\vec{x}_{k}\})Q_{q,\wbar q}(x_{k},\wbar{x}_{k})(Q_{q',\wbar q'}(x_{k},\wbar{x}_{k}))^{\ast}=\\
=(2\pi)^{4}\delta(\nu_{h}-\nu_{h}')\delta_{n_{h}n_{h}'}\delta_{\bell\bellp}\,,
\label{eq:normqq}
\end{multline}
with $|q, \wbar q\rangle$ being a product of the Baxter equation eigenstates.

\section{Baxter differential equation}
\label{sec:Bax}

Substituting the following integral ansatz \cite{Derkachov:2002wz}
\begin{equation}
Q_{q,\wbar q}(u,\wbar u)= \int\frac{d^2 z}{z\wbar z}\, 
z^{-i u} {\wbar z}^{-i\wbar u}\, \tilde Q_{q,\wbar q}(z,\wbar z)\,,
\label{eq:Q-R}
\end{equation}
to Baxter equations (\ref{eq:Baxeq})-(\ref{eq:Baxbeq})
where we integrate over the two-dimensional 
$\vec z-$plane with $\wbar z=z^*$
and $\tilde Q_{q,\wbar q}(z,\wbar z)$ depends on $\{q,\wbar q\}$
one obtains the fourth order differential equation for the function  
$\tilde Q_{q,\wbar q}(z,\wbar z)$
\begin{equation}
\left[z^s\lr{z\partial_z}^{4}z^{1-s}+z^{-s}\lr{z\partial_z}^{4}z^{s-1}
-2\lr{z\partial_z}^{4}-\sum_{k=2}^4 i^{k}q_k\lr{z\partial_z}^{4-k}
\right]\tilde Q_{q,\wbar q}(z,\wbar z)=0\,.
\label{eq:Eq-1}
\end{equation}
A similar equation holds in the anti-holomorphic sector 
with $s$ and $q_k$ replaced by $\wbar s=1-s^*$ and
$\wbar q_k=q_k^*$, respectively. 
Conditions for the analytical properties and asymptotic behaviour of 
$Q_{q,\wbar q}(u,\wbar u)$
become equivalent to a
requirement for $\tilde Q_{q,\wbar q}(z,\wbar z=z^*)$ to be a single-valued function 
on the complex $z-$plane.
The differential equation (\ref{eq:Eq-1})
possesses three regular singular points located 
at $z=0$, $z=1$ and $z=\infty$.
Around each of this points one can construct
four linearly independent solutions,
$Q_a(z)$. 
Similar relations hold for the anti-holomorphic equation
with four independent solutions  being $\wbar Q_b(\wbar z)$.

The general expression for the Baxter function reads then:
\begin{equation}
\tilde Q_{q,\wbar q}(z,\wbar z) = \sum_{a,b=1}^4 Q_a(z)\, C_{ab}\, \wbar Q_b(\wbar z)\,,
\label{eq:general-sol}
\end{equation}
where $C_{ab}$ is a mixing matrix. The functions $Q_a(z)$ and
$\wbar Q_b(\wbar z)$ 
have a nontrivial monodromy\footnote{The monodromy matrix around 
$z=0$ is defined as $Q_n^{(0)}(z\e^{2\pi i})=M_{nk}Q_k^{(0)}(z)$
and similarly for the other singular points.} 
around three singular points, 
$z,\,\wbar z=0$, $1$ and $\infty$. 
In order to be well-defined on the whole plane, 
functions $\tilde Q_{q,\wbar q}(z,\wbar z=z^*)$ 
should be single-valued and their
monodromy should cancel in the r.h.s.\ of Eq.~(\ref{eq:general-sol}). 
This requirement determines the values of the mixing coefficients,
$C_{ab}$, and also allows to calculate the quantized values of the 
conformal charges $q_k$.

It turns out that the differential equation (\ref{eq:Eq-1}) is also 
symmetric under the  transformation
$z\to 1/z$ with $q_k\to (-1)^k q_k$. 
This property leads to
\\[1mm]
\begin{equation}
\tilde Q_{q,\wbar q}(z,\wbar z)=\e^{i\theta_4(q,\wbar q)}
\tilde Q_{-q,-\wbar q}(1/z,1/\wbar z)\,,
\label{eq:Qz-symmetry}
\end{equation}
\\[0mm]
where $\pm q=(q_2,\pm q_3, q_4)$ denotes 
the integrals of motion corresponding to the function $\tilde Q_{q,\wbar q}(z,\wbar{z})$.
The above formula allows us to define the solution 
$\tilde Q_{q,\wbar q}(z,\wbar{z})$ around 
$z=\infty$ from the solution at $z=0$. Thus, applying (\ref{eq:Qz-symmetry})
we are able to find $\tilde Q_{q,\wbar q}(z,\wbar{z})$ and analytically continue it to the whole 
$z-$plane.

\subsection{Solution around $z=0$}

Solutions  $Q(z)\sim z^a$ around $z=0$ can be found by the series method.
The indicial equation for the solutions of Eq.~(\ref{eq:Eq-1})
reads as follows
\begin{equation}
(a-1+s)^4=0\,.
\label{eq:indicial-0}
\end{equation}
with four-times degenerate $a=1-s$. 
Therefore, the fundamental set of linearly independent solutions to (\ref{eq:Eq-1}) 
around $z=0$ is given by
\begin{equation}
Q_m^{(0)}(z)=z^{1-s}
\sum_{k=0}^{m-1}
\frac{(m-1)!}{k!(m-k-1)!}
u_{k+1}(z) \Log^{m-k-1}(z)\,,
\label{eq:Q-0-h}
\end{equation}
with $1\le m\le 4$. The functions 
\begin{equation}
u_m(z) = 1+\sum_{n=1}^\infty z^n\,u^{(m)}_{n}(q)\,.
\label{eq:power-series-0}
\end{equation}
are defined
inside the region $|z|<1$.
Inserting (\ref{eq:Q-0-h}) and
(\ref{eq:power-series-0}) into (\ref{eq:Eq-1}), 
one derives recurrence relations for $u^{(m)}_n(q)$.
Their explicit form can be found in the Appendix~\ref{ap:uk}.

The general
solution for $\tilde Q_{q,\wbar q}(z,\wbar z)$ around $z=0$ 
can be obtained gluing holomorphic and anti-holomorphic
sectors
\begin{equation}
\tilde Q(z,\wbar z)_{q,\wbar q} \stackrel{|z|\to 0}{=} \sum_{m,\wbar m=1}^4 Q^{(0)}_m(z)\,
C^{(0)}_{m\wbar m}\,\wbar{Q}^{(0)}_{\wbar m}(\wbar z)\,.
\label{eq:Q-0}
\end{equation}
where the 
fundamental set of solutions in the anti-holomorphic sector
can be obtained from (\ref{eq:Q-0-h}) 
by substituting $s$
and $q_k$ by $\wbar s=1-s^*$ and $\wbar q_k=q_k^*$, respectively. 
The above solution (\ref{eq:Q-0}) should be single-valued on the $z-$plane.
This implies that 
the mixing matrix $C^{(0)}_{m\wbar m}$
for $n+m\le 5$ has the following structure
\begin{equation}
C^{(0)}_{nm}=\frac{\sigma}{(n-1)!(m-1)!}
\sum _{k=0}^{4-n-m+1}{\frac {(-2)^{k}}{k!}\,\alpha_{k+n+m-1}}\,
\label{eq:C0}
\end{equation}
with $\sigma, \alpha_1,\alpha_2 ,\alpha_{3}$ being arbitrary 
complex parameters and $\alpha_4=1$. Below the main anti-diagonal,
that is for $n+m> 4+1$, $C^{(0)}_{nm}$ vanish.

Using the above formulae one can derive from (\ref{eq:HNQQ}) 
an expression for the energy 
\begin{equation}
E_4(q,\wbar q)=\Re\left[\frac{\alpha_{3}(-q,-\wbar q)}{\alpha_{4}(-q,-\wbar q)}+
\frac{\alpha_{3}(q,\wbar q)}
{\alpha_{4}(q,\wbar q)}
\right]\,,
\label{eq:E-fin}
\end{equation}
where the arbitrary complex parameters $\alpha_3$ and $\alpha_4$, 
defined in Eq.~(\ref{eq:C0}),
will be fixed by 
the quantization conditions.

\subsection{Solution around $z=1$}

Solving Eq.~(\ref{eq:Eq-1}) 
around $z=1$ with asymptotics $Q(z)\sim(z-1)^b$
we obtain the following indicial equation
\begin{equation}
(b+1+h-4s)(b+2-h-4s)(b-1)b=0\,,
\label{eq:b-exponents}
\end{equation}
where $h$ is the total $\SL(2,\mathbb{C})$ spin defined by (\ref{eq:hpar}).
Although the solutions $b=0,1$ differ 
by an integer, for $h\neq (1+n_h)/2$, 
one can construct four solutions $Q^{(1)}_i(z)$ 
without logarithmic terms.
The $\Log(1-z)-$terms are only needed for $\Im[h] = 0$ where additional 
degeneration occurs. 
The solutions with $\Im[h] \ne 0$ will be constructed in Section 
\ref{sec:traj}  while 
the ones with $\Im[h]= 0$ in Sections 
\ref{sec:tsol} and \ref{sec:psol}.

Let us consider the duality relation
(\ref{eq:Qz-symmetry}). 
Using the function $\tilde Q_{q,\wbar q}(z,\wbar{z})$  
we evaluate 
\begin{equation}
\tilde Q_{q,\wbar q}(z,\wbar z)\stackrel{|z|\to 1}{=}\sum_{m,\wbar m=1}^4
Q_m^{(1)}(z)\,C^{(1)}_{m\wbar m}\,\wbar Q_{\wbar m}^{(1)}(\wbar z)\,.
\label{eq:Q-1g}
\end{equation}
in the limit $|z|\rightarrow1$. 
In this way we obtain a set of relations for
the functions
$C^{(1)}_{m\wbar m}(q,\wbar q)$. 
The derivation is based
on the following property 
\begin{equation}
Q^{(1)}_a(1/z;-q)= \sum_{b=1}^4 S_{ab} \,Q^{(1)}_b(z;q)\,,
\label{eq:S-def}
\end{equation}
with $\Im[1/z]>0$
and where the dependence on the integrals of motion
was explicitly indicated.
Here taking $z\rightarrow 1$ limit in (\ref{eq:Q-1g}) and 
substituting to (\ref{eq:S-def}) 
we are able to
evaluate the $S-$matrix.
Similar relations hold in the anti-holomorphic sector.
The $S-$matrix does not depend on $z$
because
the $Q-$functions on both sides of relation  
(\ref{eq:S-def}) satisfy the
same differential equation (\ref{eq:Eq-1}).

Now, substituting (\ref{eq:Q-1g}) and 
(\ref{eq:S-def}) into (\ref{eq:Qz-symmetry}), we find
\\[-1mm]
\begin{equation}
C^{(1)}_{m\wbar m}(q,\wbar q) = \e^{i\theta_4(q,\wbar q)}
\sum_{n,\wbar n=1}^4 S_{nm} C^{(1)}_{n\wbar n}(-q,-\wbar q)\,
\wbar S_{\wbar n\wbar
m}\,.
\label{eq:parity-qc}
\end{equation}
Using the mixing matrix at $z=1$
we can compute from Eq.~(\ref{eq:parity-qc}), 
the eigenvalues of the quasimomentum, $\theta_4(q,\wbar q)$.

\subsection{Transition matrices}

In the previous sections we discussed
the solutions $\tilde Q_{q,\wbar q}(z,\wbar z)$ to (\ref{eq:Eq-1}) in
the vicinity of $z=0$ and $z=1$.
In order to calculate 
$\tilde Q_{q,\wbar q}(z,\wbar z)$ on the whole complex $z-$plane 
 one can glue these solutions in the region
$|1-z|<1,\,|z| < 1$ and, then analytically continue the
resulting expression for $\tilde Q_{q,\wbar q}(z,\wbar z)$ 
by making use of the duality relation (\ref{eq:Qz-symmetry}).

To this end, we define
the transition matrices $\Omega(q)$ and $\wbar \Omega(\wbar q)$:
\begin{equation}
Q_n^{(0)}(z)=\sum_{m=1}^4 \Omega_{nm}(q)\, Q_m^{(1)}(z)\,,\qquad
\wbar Q_{n}^{(0)}(\wbar z)=\sum_{m=1}^4\wbar \Omega_{nm}(\wbar q)
 \,\wbar Q_{m}^{(1)}(\wbar z)\,.
\label{eq:Omega-def}
\end{equation}
which are uniquely fixed \cite{Derkachov:2002wz}
and do not depend on $z$ and $\wbar z$.
Substituting 
(\ref{eq:Omega-def}) into
(\ref{eq:Q-0}) and matching the result into 
(\ref{eq:Q-1g}), we find 
the following
relation
\begin{equation}
C^{(1)}(q,\wbar q)=\left[\Omega(q)\right]^T C^{(0)}(q,\wbar q)\ \wbar
\Omega(\wbar{q})\,.
\label{eq:C1-C0}
\end{equation}
The above matrix formula contains
$16$ complex equations with:
three $\alpha-$pa\-ram\-e\-ters inside the matrix $C^{(0)}$,
six parameters\footnote{Here, this number
is taken for ordinary solutions but it depends on values of 
conformal charges and will be calculated in the next sections.} 
inside the matrix $C^{(1)}$,
two integrals of motion $q_3 ,q_4$ with $\wbar
q_k=q_k^*$, where $q_2=q_2^*$ has been already quantized (\ref{eq:hpar}).
Thus, one obtains five nontrivial consistency conditions.
Next, Eq.~(\ref{eq:C1-C0}) 
allows us to determine the matrices $C^{(0)}$ and $C^{(1)}$ and 
provides the quantization conditions for the integrals of
motion, $q_k$ and $\wbar q_k$ with $k=3,4$. 
The numerical solutions \cite{Kotanski:2001iq} 
to the quantization conditions (\ref{eq:C1-C0}) 
will be presented in details in the next sections.
Finally, using these quantized conformal charges one can evaluate 
the eigenvalues of the Baxter $\mathbb{Q}-$operator (\ref{eq:Q-R}) and
the Hamiltonian eigenvalues (\ref{eq:E-fin}).

\subsection{Norm with integral ansatz}

Substituting the ansatz (\ref{eq:Q-R}) to (\ref{eq:normqq})
we can also rewrite the norm  in terms of $\tilde{Q}_{q,\wbar q}(z,\wbar{z})$
solutions.
To evaluate this norm we use the following
correspondence
\begin{equation}
\nonumber
xQ_{q,\wbar q}(x,\wbar{x}) = -i\int d^{2}z\  z^{-ix-1}\wbar{z}^{-i\wbar{x}-1}z\ \partial_{z}\ \tilde{Q}_{q,\wbar q}(z,\wbar{z})\,,
\end{equation}
with 
$n=x-\wbar{x}$, $2\lambda=x+\wbar{x}$ and
\begin{equation}
\int_{-\infty}^{\infty}d\lambda\sum_{n=-\infty}^{\infty}\  z^{-ix}\wbar{z}^{-i\wbar{x}}\  w^{ix}\wbar{w}^{i\wbar{x}}=2\pi^{2}z\wbar{z}\delta^{(2)}(z-w)\,.
\end{equation}
The final expression for 
the scalar product defined in (\ref{eq:normQQ}) reads as
\begin{multline}
\langle q,\wbar q|q',\wbar q'\rangle=\frac{4^{4}}{6 \pi^{6}}
\left(\prod_{k=1}^{3}\frac{d^{2}z_{k}}{z_{k}\wbar{z}_{k}}\right)
\left(\tilde{Q}_{q',\wbar q'}(z_{1},\wbar{z}_{1})\ldots\tilde{Q}_{q',\wbar q'}
(z_{3},\wbar{z}_{3})\right)^{\ast}\\
\times\left(\prod_{1\le i<j}^{3}(-D_{ij}\wbar{D}_{ij})\right)
\tilde{Q}_{q,\wbar q}(z_{1},\wbar{z}_{1})\ldots
\tilde{Q}_{q,\wbar q}(z_{3},\wbar{z}_{3})\:,
\label{eq:normzz}
\end{multline}
where $D_{ij}=z_{i}\partial_{i}-z_{j}\partial_{j}$.
In this work the above expression is used to test the normalization condition.

\section{Trajectory-like solutions and their analytical continuation}
\label{sec:traj}

In this section we parameterize possible 
values of conformal charges with use of WKB approximation.
Next, we construct trajectory-like solutions with $\nu_h \in \mathbb{R}$
and perform their analytical continuation into $\nu_h \in \mathbb{C}$
that allows to explain the existence of specific point-like solutions with $\nu_h=0$.

\subsection{WKB approximation}
Finding asymptotic solutions \cite{Derkachov:2002pb} 
to the Baxter equations, (\ref{eq:Baxeq})
and (\ref{eq:Baxbeq}),
one can derive the WKB approximation of
the conformal charges $q_3$ and $q_4$.

In the limit of large values of conformal charges, 
\ie where $q_4^{1/4}\gg q_3^{1/3} \gg q_2^{1/2}$
\begin{equation}
q_4^{1/4}=\frac{\Gamma^2(3/4)}{4\sqrt{\pi}}\left[ \frac1{\sqrt {2}}
\ell_1+
\frac{i}{\sqrt {2}}\ell_2 \right]\,.
\label{eq:q4-quan}
\end{equation}
Thus, for fixed $q_2$ and $q_3$ 
the quantized values of $q_4^{1/4}$ can be identified
with $(\ell_1,\ell_2)$-vertices of square-like WKB lattices.
One can see an example of such a correspondence in Fig.~\ref{fig:pln2}.
In the following the $q_k-$spectrum related to vertices of WKB
lattice will be called lattice spectrum or $q_k-$lattice.

The next two integers, $\ell_3$ and $\ell_4$, establish
dependence between values of $q_4$ and $q_3$ 
\cite{Kotanski:2005ci,Kotanski:2006sh}.
The leading approximation 
for the charge $q_3$ reads
\begin{equation}
\Im \left[\frac{q_3}{q_4^{1/2}}\right] 
=\ell_3,
\label{eq:q3-quan}
\end{equation}
where $\ell_3$ describes a ``twist'' between  $q_4^{1/4}$ and 
$q_3^{1/2}$ WKB lattices. For $\ell_3=0$ one obtains resemblant lattices
without any ``twist''.
An example of such spectrum is plotted in Fig.~\ref{fig:l43s01n1}
where the $q_4^{1/4}-$lattice has similar structure to
the $q_3^{1/2}-$lattice.
For $\ell_3 \ne 0$ $q_4^{1/4}-$lattice winds
faster that corresponding $q_3^{1/2}-$lattice.
This situation is presented in Fig.~\ref{fig:l43w01n1}.

The system of Eqs.~(\ref{eq:q4-quan}) and (\ref{eq:q3-quan}) 
is underdetermined and 
it does not fix the charge $q_3$ completely.
The last parameter, $\ell_4$, describes  the scale between $q_4^{1/4}$ 
and $q_3^{1/2}$ WKB lattices 
but its exact analytical 
relation to conformal charges even in the WKB approximation is unknown.

The quasimomentum in the WKB approximation reads as follows
\begin{equation}
\theta_4=-\frac{\pi}2\ell_{\theta}
=\frac{\pi}2 
(\ell_2+\ell_3-\ell_1)\qquad ({\rm mod}~ 2\pi)\,,
\label{eq:theta-4}
\end{equation}
where $\ell_1$, $\ell_2$ and $\ell_3$ are even for even $\ell_{\theta}$ and
odd for odd $\ell_{\theta}$.
Thus, we have two kinds of lattices: with $\theta_4=0,\pi$ and
with $\theta_4= \pi/2, 3\pi/2$.
Therefore, for a given lattice not all vertices are occupied.

The last conformal charge, $q_2=-h(h-1)$, depends on $n_h$ and $\nu_h$
(\ref{eq:hpar}). Since, $\nu_h$ is real and continuous
the conformal charges form one dimensional trajectories.
One can see 
an example of the trajectories
in Figs.~\ref{fig:q4traj} and \ref{fig:entraj}.
The WKB approximation applies for large values of
conformal charges
therefore our exact numerical results agree better with WKB
for $q_4 \to \infty$.
When $q_4$ goes to zero the approximation stops to work.
The numerical results from Ref.~\cite{Kotanski:2006sh}
and from the present paper
show lack of
quantized values of $q_4^{1/4}$
in the centre of the lattice,
for example in Fig.~\ref{fig:pln2}.

\subsection{Solution around $z=1$ for $\Im[h]\neq 0$}

In this Section we briefly recapitulate the method to solve
Eq.~(\ref{eq:Eq-1}) exactly. For $\nu_h\neq 0$ 
and $n_h>0$
we define around $z=1$
the  following fundamental set of solutions\footnote{Exchange
$n_h \to -n_h$ corresponds to $Q_3^{(1)}(z) \leftrightarrow Q_4^{(1)}(z)$}
\begin{eqnarray}
\nonumber
&&Q_m^{(1)}(z) = z^{1-s} (1-z)^{m-1} v_m(z)\,,
\\[2mm]
&&Q_{3}^{(1)}(z) = z^{1-s} (1-z)^{4s-h-1}v_{3}(z)\,,
\nonumber
\\[2mm]
&&Q_{4}^{(1)}(z) = z^{1-s} (1-z)^{4s+h-2}v_{4}(z)\,,
\label{eq:set-1}
\end{eqnarray}
where $m=1,2$. 
The functions $v_k(z)$ are given by the
power series
\begin{equation}
v_k(z)=\sum_{n=0}^\infty (1-z)^n \,v^{(k)}_{n}(q)\,,
\label{eq:v-series}
\end{equation}
which converge inside the region $|1-z|<1$
and where 
the expansion coefficients
 $v^{(k)}_{n}$ satisfy the four-term 
recurrence relations\footnote{
The factor $z^{1-s}$ was included in the r.h.s.\ of 
(\ref{eq:set-1}) 
to simplify the form of the recurrence relations.}
with respect to the index $n$.
Their explicit form can be found in the Appendix~\ref{ap:vk}.
Similar calculations have to be performed in the anti-holomorphic sector
with $s$ and $h$ replaced by
$\wbar s=1-s^*$ and $\wbar h=1-h^*$, respectively. 

\begin{figure}[ht!]
\centerline{
\begin{picture}(160,52)
\put(26,2){\epsfysize5.1cm \epsfbox{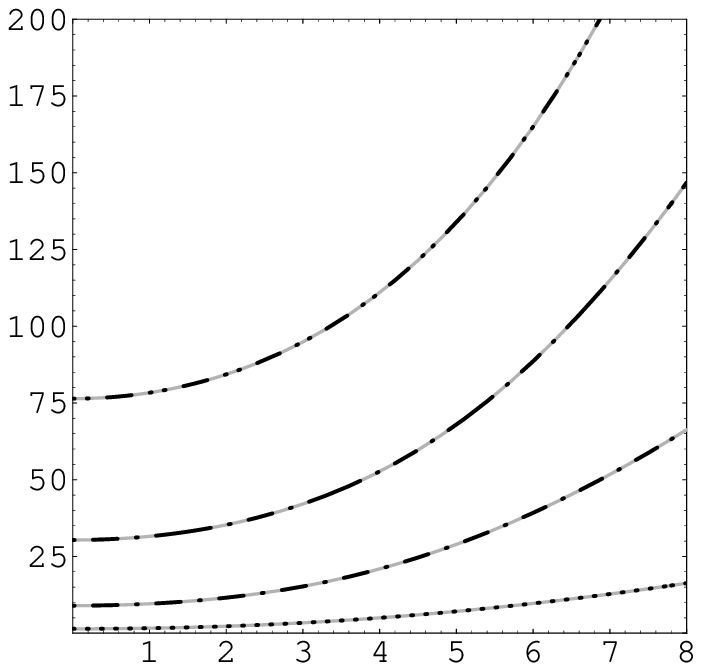}}
\put(89,2){\epsfysize5.cm \epsfbox{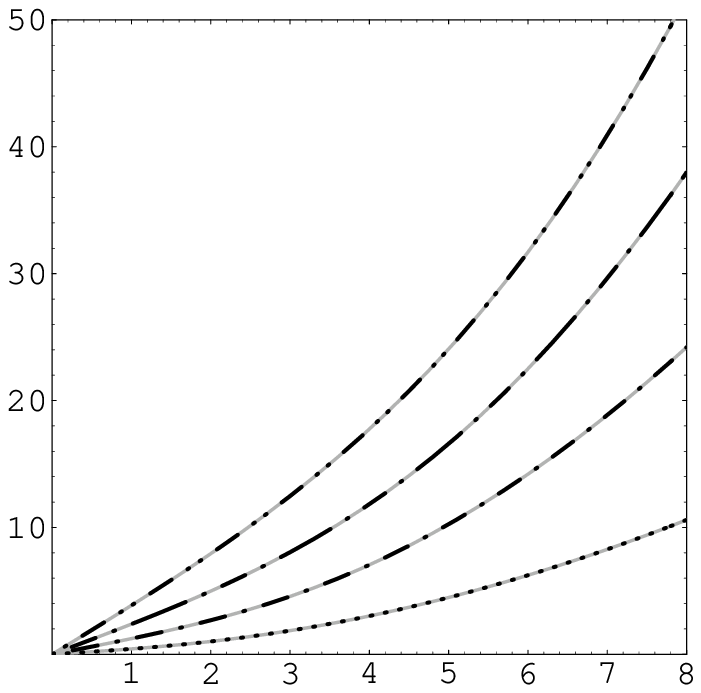}}
\put(69,0){$\nu_h$}
\put(22,27){\rotatebox{90}{$\Re[q_4]$}}
\put(131,0){$\nu_h$}
\put(85,27){\rotatebox{90}{$\Im[q_4]$}}
\end{picture}
}
\caption{Trajectories of conformal charges for $q_3=0$ and 
$h=\frac3{2}+i \nu_h$ where $\nu_h \in \mathbb{R}$}
\label{fig:q4traj}
\end{figure}

\begin{figure}[ht!]
\centerline{
\begin{picture}(90,41)
\put(10,0){\epsfysize4.5cm \epsfbox{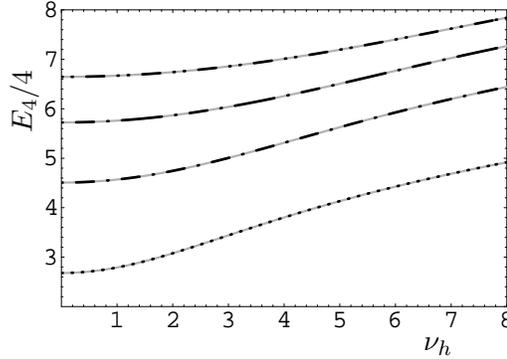}}
\put(63,-2){$\nu_h$}
\put(8,27){\rotatebox{90}{$E_4/4$}}
\end{picture}
}
\caption{The energy along the trajectories from Fig.~\ref{fig:q4traj}
as a function of $\nu_h$}
\label{fig:entraj}
\end{figure}

A general solution for $\tilde Q_{q,\wbar q}(z,\wbar{z})$ for $\Im[h] \ne 0$
that satisfies
the singlevaluedness condition
can be constructed as 
\begin{equation}
\small
\tilde Q_{q,\wbar q}(z,\wbar z)\stackrel{|z|\to 1}{=}\beta_h Q_4^{(1)}(z)\wbar Q_4^{(1)}(\wbar
z)+\beta_{1-h} Q_3^{(1)}(z)\wbar Q_3^{(1)}(\wbar z) +\sum_{m,\wbar m=1}^{2}
Q_m^{(1)}(z)\,C^{(1)}_{m\wbar m}\,\wbar Q_{\wbar m}^{(1)}(\wbar z)\,.
\label{eq:Q-1}
\end{equation}
Here
the parameters $\beta_h=C^{(1)}_{44}$
while $\beta_{1-h}=C^{(1)}_{33}$.
The $\beta-$coefficients depend, in general, on the total spin $h$ and
$\wbar h=1-h^*$. They are chosen 
in such a way that the symmetry of
the eigenvalues of the Baxter operator under $h\to 1-h$ becomes manifest. 
Thus, the mixing matrix $C^{(1)}$ defined in (\ref{eq:Q-1}) 
depends on $2+2^2$ complex parameters $\beta_h$, $\beta_{1-h}$
and $C^{(1)}_{m\wbar m}$ which are functions 
of the integrals of
motion $(q,\wbar q)$, so they can be fixed by the quantization conditions. 

The functions defined by Eqs.~(\ref{eq:set-1}) and (\ref{eq:Q-0-h}) 
give the solutions $\tilde Q_{q,\wbar q}(z,\wbar z)$.
Their conformal charges calculated from Eq.~(\ref{eq:C1-C0})
form in the $q_k-$space continuous trajectories
propagating in the $\nu_h-$real space.
An example of such trajectories 
for $q_3=0$ and 
$h=\frac3{2}+i \nu_h$ where $\nu_h \in \mathbb{R}$
is plotted in Fig.~\ref{fig:q4traj}.
We can also calculate the energy along the trajectories,
Fig.~\ref{fig:entraj}.
However, for $\nu_h=0$ solutions to Eq.~(\ref{eq:Eq-1}) have a different form. 
Therefore, one has to consider these solutions separately. 
We will discuss them in the next sections.

\subsection{Analytical continuation}

Let us consider analytical continuation of $E_4(\nu_h)$
into the complex $\nu_h-$ plane, the same as the one performed in
calculation of anomalous dimensions in deep inelastic scattering 
processes \cite{Korchemsky:2003rc}. 
The requirement for $n_h$ to be integer follows from condition for 
singlevaluedness of the wave-function $\Psi_{\vec{p},\{q,\wbar q\}}$.
On the other hand $\Psi_{\vec{p},\{q,\wbar q\}}$ is normalizable
with the respect to Eq.~(\ref{eq:norm}) when $\nu_h$ is real.
Therefore performing analytical continuation we relax the 
normalization condition
(\ref{eq:normqq}) while keeping the quantization conditions (\ref{eq:C1-C0}).
This in turn implies that 
the operator of Hermitian conjugation,
which is related to the scalar product,
is not well defined. 
It follows that:
\begin{equation}
h^\ast \ne 1- \wbar h\,, \qquad q_k^\ast \ne \wbar q_k\,.
\label{eq:hhbqqb}
\end{equation}

\begin{figure}[ht]
\centerline{
\begin{picture}(150,40)
\put(10,2){
\epsfysize4.0cm \epsfbox{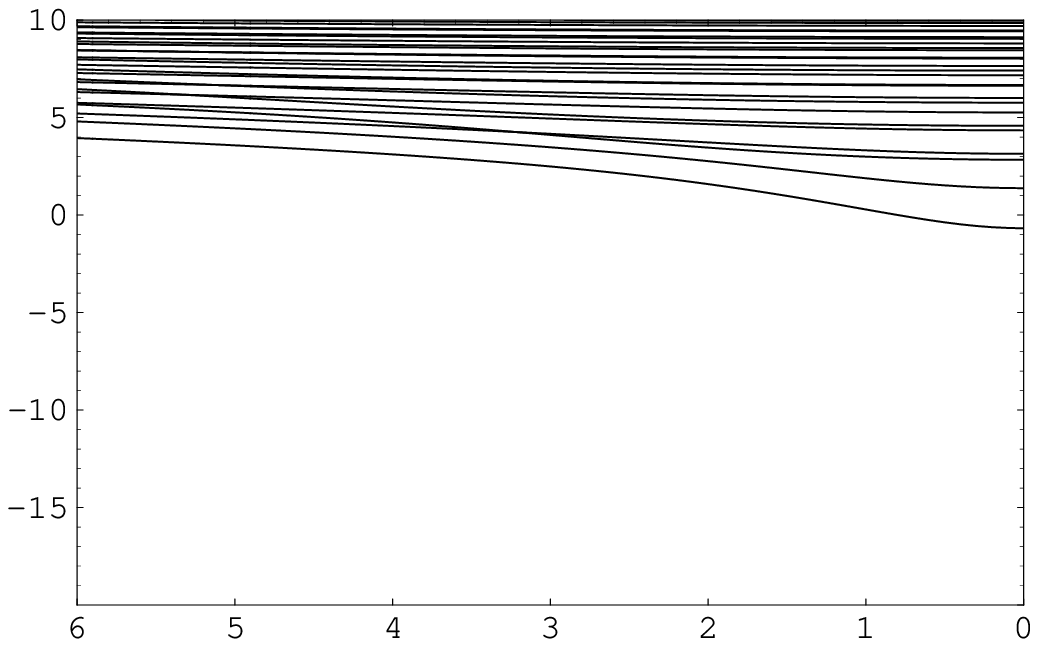}}
\put(76,2){\epsfysize4.0cm \epsfbox{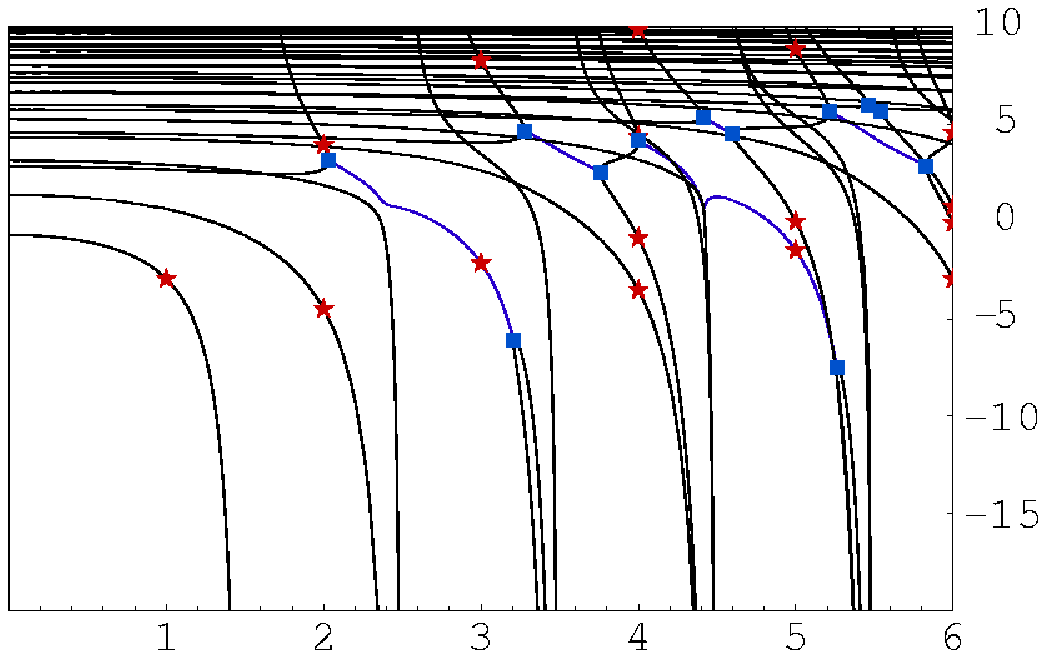}}
\put(18,-1){$ \nu_h^{(a)}$}
\put(127,-1){$i \nu_h^{(a)}$}
\put(10,20){\rotatebox{90}{$\Re [E_4/4]$}}
\end{picture}
}
\caption{Energy and its analytical continuation  in  
$h^{(a)}=\frac1{2}+i \nu_h^{(a)} \in \mathbb{R}$ direction. 
Here, {\em stars} 
are related to the twin-solutions of
the point-like solutions with $n_h^{(p)}\in 2 \mathbb{Z}$.
For $q_3=0$ and  $n_h^{(p)}\in 2 \mathbb{Z}+1$ we have only point-like 
solutions.Trajectories between branching points 
({\em boxes}) have non-vanishing 
imaginary part of $E_4$ and $q_4=\wbar q_4$.}
\label{fig:Z2E}
\end{figure}

One can notice that in the analytical continuation region
there is a transformation
\begin{equation}
(h,\wbar h) \to (h,1-\wbar h)\,,
\label{eq:hhsym}
\end{equation}
which does not change the values of Casimirs (\ref{eq:evq2}) 
and the quantization conditions are also invariant under this
transformation\footnote{Due to the symmetry
of the spectrum $(h,\wbar h)\to(1-h,1-\wbar h)$, 
we have also a transformation
$(h,\wbar h) \to (1-h,\wbar h)$,
which is analogical to (\ref{eq:hhsym}).}.
The symmetry (\ref{eq:hhsym}) relates two different solutions from
analytically continued sheets
which have the same conformal charges but
different conformal weights, \ie $h^{(p)}$ and $h^{(a)}$.
Requirements $h^{(p)}=h^{(a)}$ and $\wbar h^{(p)}=1- \wbar h^{(a)}$
give conditions 
\begin{equation}
\Re[\nu_h^{(a)}]=\Re[\nu_h^{(p)}]=0\,,\quad
n_h^{(a)}=- 2 \Im [\nu^{(p)}_h]\,, \quad 
n_h^{(p)}=- 2 \Im [\nu^{(a)}_h]\,. 
\label{eq:symcon}
\end{equation}
Since conformal Lorentz spins are integer,
the symmetry (\ref{eq:hhsym}) exists only 
for half-integer and integer conformal weights.
Now, if $h^{(p)}$ has  $\nu_h^{(p)} \in \mathbb{R}$ 
and another one, $h^{(a)}$, from analytical continuation
region has $\nu_h^{(a)} \in \mathbb{C}$
then Eq.~(\ref{eq:symcon}) implies that
$n_h^{(a)}=0$, $\nu_h^{(p)}=0$ and $n_h^{(p)}=2 i \nu_h^{(a)}$.
Thus, 
all solutions
with $h^{(a)}=\wbar h^{(a)}=\frac{1}{2}+i \nu_h^{(a)} \in \mathbb{Z}/2$,
\ie situated on analytical continuation
(Riemann) surfaces of  $n^{(a)}_h=0$ physical states,
have their twin-solutions with 
$h^{(p)}=\frac{1+n_h^{(p)}}{2},\; 
\wbar h^{(p)}=\frac{1-n_h^{(p)}}{2} \in \mathbb{Z}/2$, 
where $n_h^{(p)}/2=i\nu_h^{(a)}$.
The symmetry (\ref{eq:hhsym}) can move the twin-solutions
outside the physical region therefore solutions denoted
by $(p)$ can be either normalizable or non-normalizable.

Let us focus on the set of solutions with $q_4=\wbar q_4$ 
and $q_3=\wbar q_3=0$.
On the left panel of Fig.~\ref{fig:Z2E}  energy of such trajectories with 
$n_h=0$ and $\nu_h \in \mathbb{R}$ is plotted.
Relaxing the normalization condition (\ref{eq:norm})
we perform analytical continuation of this energy plot 
to complex $\nu_h-$plane.
The energies as  functions of $\nu_h \in \mathbb{C}$
form complicated Riemann surfaces.
One can see their cross-section  in $i \nu_h \in \mathbb{R}$ 
direction on the right panel
of Fig.~\ref{fig:Z2E}.
For example, due to the symmetry (\ref{eq:hhsym}) 
and the conditions (\ref{eq:symcon}),
the solutions for $i \nu_h^{(a)}=1$
should have twin-solutions with $n_h^{(p)}=2$
and $\nu_h^{(p)}=0$. Performing numerical calculation
one can find physical trajectories for all solutions with 
$h^{(p)}=\frac{3}{2}+i \nu_h^{(p)}$,
\eg the ones from Figs.~\ref{fig:q4traj} and \ref{fig:entraj},
except the twin-solution to the solution
with $i \nu_h^{(a)}=1$ 
denoted in Fig.~\ref{fig:Z2E} by {\em star}. 
It means that the latter twin-solution is not situated on any trajectory,
\ie it is a point-like solution.
Other similar point-like twin-solutions 
without physical trajectories 
exist to all solutions 
denoted in this figure by {\em stars}.
Moreover, since 
there is not any trajectory
for $q_4=\wbar q_4$ and $q_3=\wbar q_3 =0$
all solutions with $i \nu_h \in \frac{2 \mathbb{Z}+1}{2}$ 
from the right panel of Fig.~\ref{fig:Z2E}
have their point-like twin-solutions.

The point-like solutions are candidates for Hamiltonian eigenstates 
(\ref{eq:Schr}) with discrete spectrum.
The point-like states do not depend on continuous 
parameter $\nu_h$.
In order to contribute to the decomposition of the Hamiltonian
over the eigenstates, their scalar product (\ref{eq:normzz})
should be proportional to 
$\delta_{n_{h}n_{h}'}\delta_{\bell\bellp}$
without $\delta(\nu_h-\nu_h')$.
On the other hand the eigenstates should properly transform
with respect to $\SL(2,\mathbb{C})$ group.
This implies that the scalar product (\ref{eq:normzz}) 
should be infinite or equal to zero.
Thus, one may suppose
that the point-like solutions are non-normalizable with respect
to the $\SL(2,\mathbb{C})$ scalar product (\ref{eq:norm}).
In the next sections 
we present properties and check normalizability 
of the solutions with half-integer and integer conformal weights
in more detail.

\section{Solution from the trajectories with $\nu_h=0$}
\label{sec:tsol}

Let us consider an ordinary
solution\footnote{A solution which has a common structure 
for almost all values of conformal charges. 
Contrary to this case,
for special 
values of conformal charges there are other special solutions 
with smaller number of $\Log(1-z)-$terms.} 
to Eq.~(\ref{eq:Eq-1})
around $z=1$ with $\nu_h=0$.
Conformal charges of these ordinary 
solutions are situated on $q_k-$trajectories propagating in 
continuous $\nu_h$.
Thus, these solutions should be normalized to Dirac delta function.
Indeed, one can find by explicit calculation that their scalar product
is proportional to $\delta(\nu_h-\nu_h')$.
In this section we are going to describe construction
of the ordinary solutions whose
conformal charges form resemblant, winding 
and descendent lattices (\ref{eq:q4-quan})-(\ref{eq:q3-quan}).

\subsection{Trajectory-like solutions around $z=1$ for even $n_h$}

Solving Baxter differential equations
with $n_h \in 2 \mathbb{Z}_+$ and $\nu_h=0$
one constructs
\begin{equation}
Q_i^{(1)}(z)=z^{1-s} v_i(1-z)\,,
\label{eq:Qnzs}
\end{equation}
where
\begin{eqnarray}
&&v_m(y) = y^{b_m} \sum_{n=0}^{\infty} v_n^{(m)} y^n\,,
\nonumber
\\[2mm]
&&v_4(y) = y^{b_4} \sum_{n=0}^{\infty} v_n^{(4)} y^n+ \Log(y) v_2(y)\,,
\label{eq:v1q3s1a}
\end{eqnarray}
with $m=1,2,3$ and 
$b_1=1$, $b_2=0$, 
$b_3=h-2+4s$, 
$b_4=-h-1+4s$.
The coefficients $v_k$ 
satisfy
the four-term recurrence relations with the boundary condition 
$v_{n_h}=1$. Similar solutions can be found in the
anti-holomorphic sector.

Constructing the general solution (\ref{eq:Q-1g})
for $h=(1+n_h)/2$ the two terms on the
r.h.s.\ of (\ref{eq:Q-1}) look differently in virtue of (\ref{eq:v1q3s1a})
\begin{equation}
\tilde Q_{q,\wbar q}(z,\wbar z)
=\beta_1\! \left[Q_4^{(1)}(z)\wbar
Q_3^{(1)}(\wbar z)+Q_3^{(1)}(z)\wbar Q_4^{(1)}(\wbar z)\right]+\beta_2\,
Q_3^{(1)}(z)\wbar Q_3^{(1)}(\wbar z) + \ldots \, ,
\end{equation}
where ellipses denote the remaining terms. Substituting 
(\ref{eq:Q-1}) into (\ref{eq:Q-R})
and performing integration over the region of $|1-z|\ll 1$, one can find the
asymptotic behaviour of $Q(u,\wbar u)$ at large $u$.

\begin{figure}[ht]
\centerline{
\begin{picture}(140,57)
\put(10,2){\epsfysize5.5cm \epsfbox{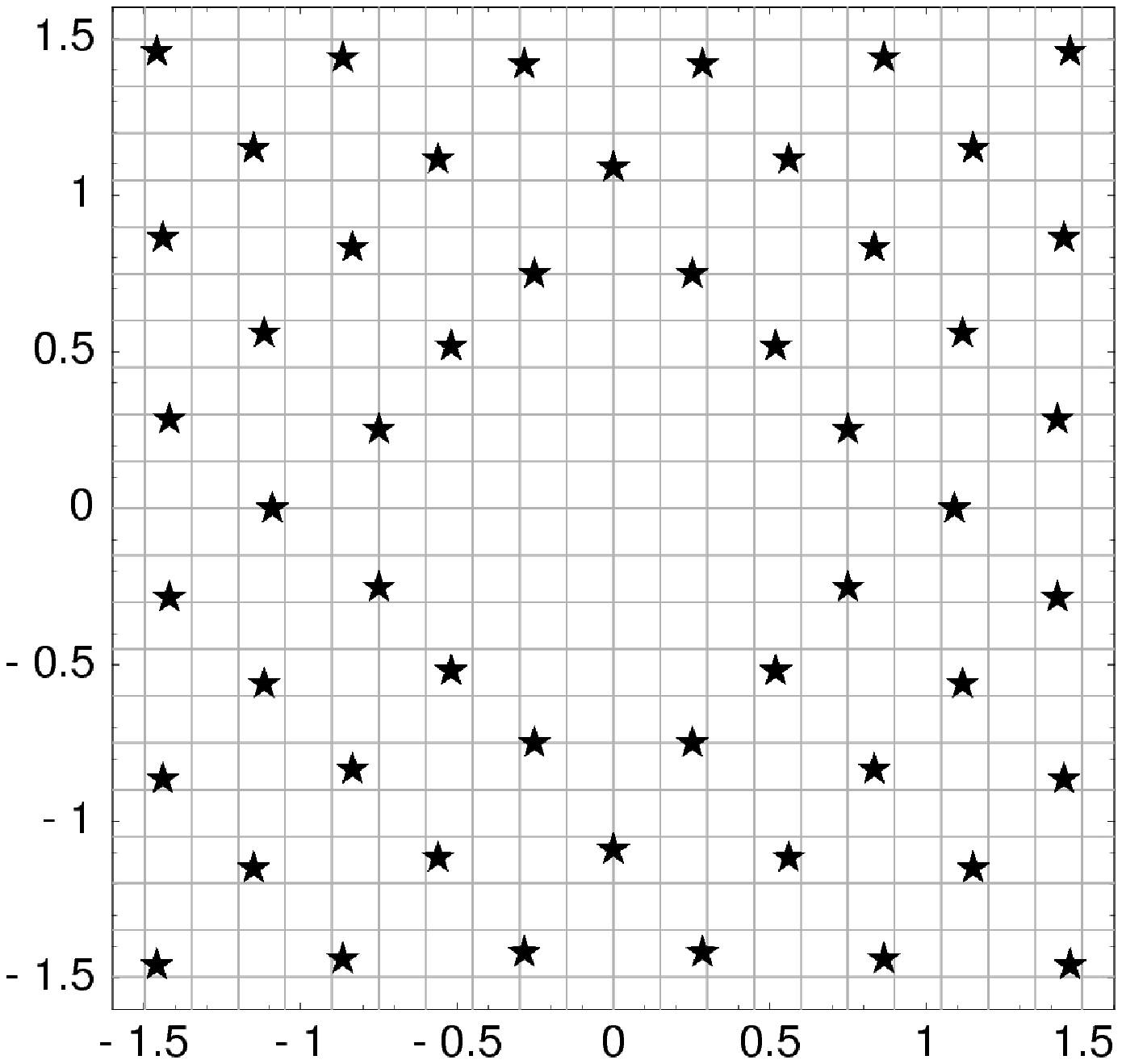}}
\put(35,-1){$\Re[q_4^{1/4}]$}
\put(7,27){\rotatebox{90}{$\Im[q_4^{1/4}]$}}
\put(70,30){
\begin{tabular}{|c|c|}
\hline
$q_4$ 	& $E_4/4$ \\
\hline \hline
$-0.289249 $&$0.651412 $ \\ \hline
$0.104882 + 0.376338 i $&$1.49514 $ \\ \hline
$1.41003 $&$ 2.67956$ \\ \hline
$-1.93065 $&$ 2.92734$ \\ \hline
$-0.69567+2.3275 i $&$3.19451 $ \\ \hline
$3.07957+3.12041 i $&$3.79614 $ \\ \hline
$-6.98331 $&$4.24321 $ \\ \hline
$-4.44222-6.59389 i $&$4.37829 $ \\ \hline
$8.9601 $&$4.50774 $ \\ \hline
\end{tabular}
}
\end{picture}
}
\caption{The numerical spectrum of quantized $q_4^{1/4}$ for 
trajectory-like solutions with $n_h=2$ and $q_3=0$ denoted by {\em stars}
on the background of the WKB lattice. 
According to 
Eq.~(\ref{eq:q3-quan}) and Eq.~(\ref{eq:theta-4}) only some vertices 
of WKB lattice are occupied.}
\label{fig:pln2}
\end{figure}


Finally, solving 
the quantization conditions
(\ref{eq:C1-C0}) one can find that conformal charges
of (\ref{eq:Qnzs}) are situated on continuous trajectories.
Spectrum of conformal charges for $n_h=0$ 
in Refs.~\cite{Kotanski:2005ci,Kotanski:2006sh}
as well as $n_h \ne 0$ can be described by WKB approximation
(\ref{eq:q4-quan})-(\ref{eq:theta-4})
where parameters $\ell_i$ with $i=1,\ldots,4$ enumerate
vertices of WKB lattices.
For example in Fig.~\ref{fig:pln2} one can see such
lattice spectrum with $q_3=0$ and $\nu_h=0$
whose continuation in $\nu_h \in \mathbb{R}$ form trajectories
depicted 
in Fig.~\ref{fig:q4traj}.
It turns out that the solutions with the lowest energy 
belong to the $n_h=0$ sector \cite{Korchemsky:2001nx}.
All other states, also those with $n_h \ne 0$, have  larger energy.

\subsection{Trajectory-like 
solutions around $z=1$ for odd $n_h$ }

Since 
for the ordinary solutions 
with $h=(1+n_h)/2 \in \mathbb{Z}$, \ie $n_h$ odd,
all roots of Eq.~(\ref{eq:b-exponents}) are integer
we have two additional terms:
$\Log(1-z)$ and $\Log^2(1-z)$.
Here, the conformal charges form similar lattice 
structures to the previous $n_h$ even case.
The states also are situated on trajectories with  $\nu_h\ne 0$
defined by Eq.~(\ref{eq:set-1}).
For instance in the $(h,\wbar h)=(1,0)$ sector with $q_3\ne 0$
these solutions have a form 
$Q_i^{(1)}(z)=z^{1-s} v_i(1-z)$ where
\begin{eqnarray}
&&v_m(y) = y^{b_m} \sum_{n=0}^{\infty} v_n^{(m)} y^n\,,
\nonumber
\\[2mm]
&&v_3(y) = y^{b_3} \sum_{n=0}^{\infty} v_n^{(3)} y^n+ \Log(y) v_2(y)\,,
\nonumber
\\[2mm]
&&v_4(y) = y^{b_4} \sum_{n=0}^{\infty} v_n^{(4)} y^n+2 \Log(y)
y^{b_3} \sum_{n=0}^{\infty} v_n^{(3)} y^n+
\Log^2(y) v_2(y)\,,
\label{eq:v1q3s1b}
\end{eqnarray}
while $m=1,2$ and $b_1=1$, $b_2=0$, $b_3=-1$, $b_4=-2$ for $s=0$
and  $b_1=0$, $b_2=3$, $b_3=2$, $b_4=1$  for $s=1$.

\begin{figure}[ht!]
\centerline{
\begin{picture}(150,57)
\put(15,2){\epsfysize5.5cm \epsfbox{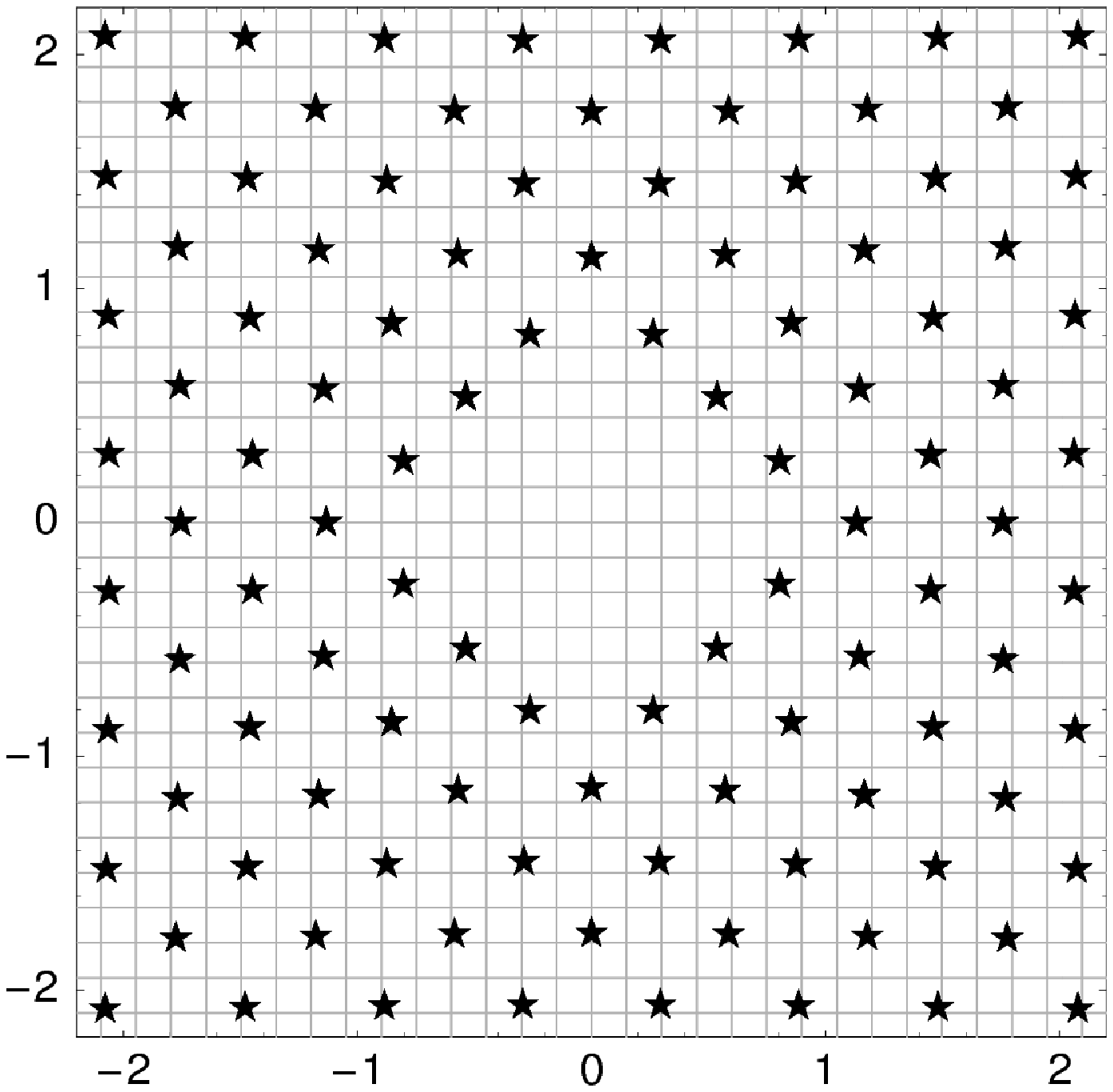}}
\put(77,1){\epsfysize5.7cm \epsfbox{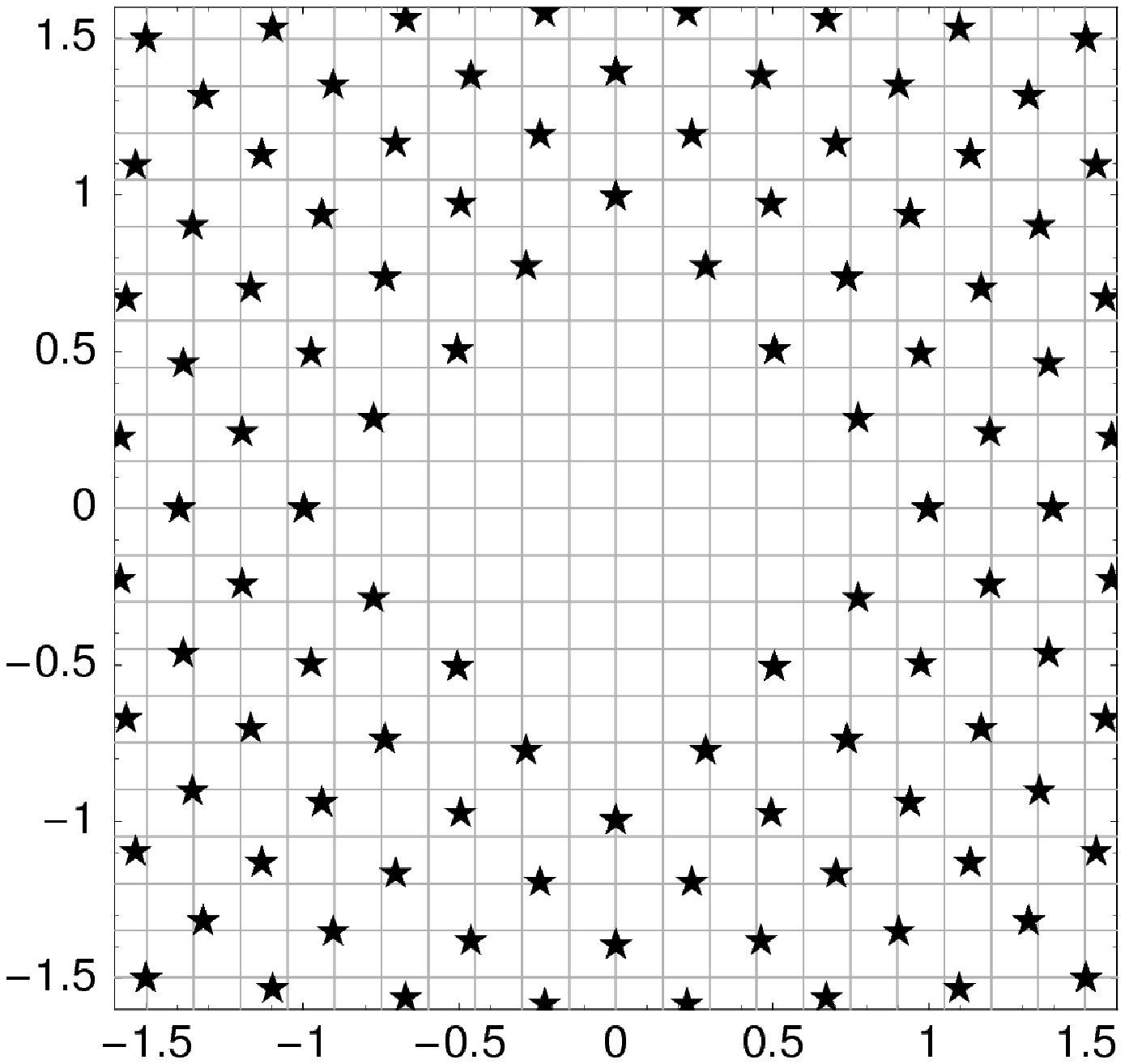}}
\put(40,-2){$\Re[q_4^{1/4}]$}
\put(11,27){\rotatebox{90}{$\Im[q_4^{1/4}]$}}
\put(104,-2){$\Re[q_3^{1/2}]$}
\put(74,27){\rotatebox{90}{$\Im[q_3^{1/2}]$}}
\end{picture}
}
\caption{The resemblant spectra of 
the conformal charges for $n_h=1$ with $\theta_4=0$. 
On the left panel the spectrum of $q_4^{1/4}$, while on the right
panel the spectrum of $q_3^{1/2}$}
\label{fig:l43s01n1}
\end{figure}

\begin{table}[ht!]
\begin{center}
\begin{tabular}{|c|c|c|}
\hline
$q_3$ 	& $q_4$ 	& $E_4/4$ \\
\hline \hline
$0.512164 i	$&$  -0.332886 $&$1.10034 $ \\ \hline
$	-0.516335 + 0.443489 i $&$ 0.154318 - 0.489842 i  $&$1.72876 $ \\ \hline
$	-0.991547$&$1.65225   $&$2.79015 $ \\ \hline
$1.0874 i	$&$ -2.12728  $&$3.06668  $ \\ \hline
$-0.70143 + 0.965223i	$&$ -0.739773 - 2.58049 i $&$3.30279  $ \\ \hline
$1.36576 + 0.578827	i$&$ 3.36485 + 3.37562 i $&$3.86627  $ \\ \hline
$1.76288i	$&$ -7.38429  $&$4.31192  $ \\ \hline
$0.865622 + 1.64001	i$&$-4.66799 + 6.98328 i  $&$4.43945  $ \\ \hline
$-1.94374$&$ 9.50289  $&$ 4.55665 $ \\ \hline
$-1.69376 + 1.27953 i$&$ 3.32987 - 11.3531 i $&$4.77895  $ \\ \hline
\end{tabular}
\end{center}
\caption{Resemblant lattice state spectrum for  $n_h=1$ and $\theta_4=0$.}
\label{tab:l43s01n1}
\end{table}

\begin{figure}[ht!]
\centerline{
\begin{picture}(140,57)
\put(12,2){\epsfysize5.5cm \epsfbox{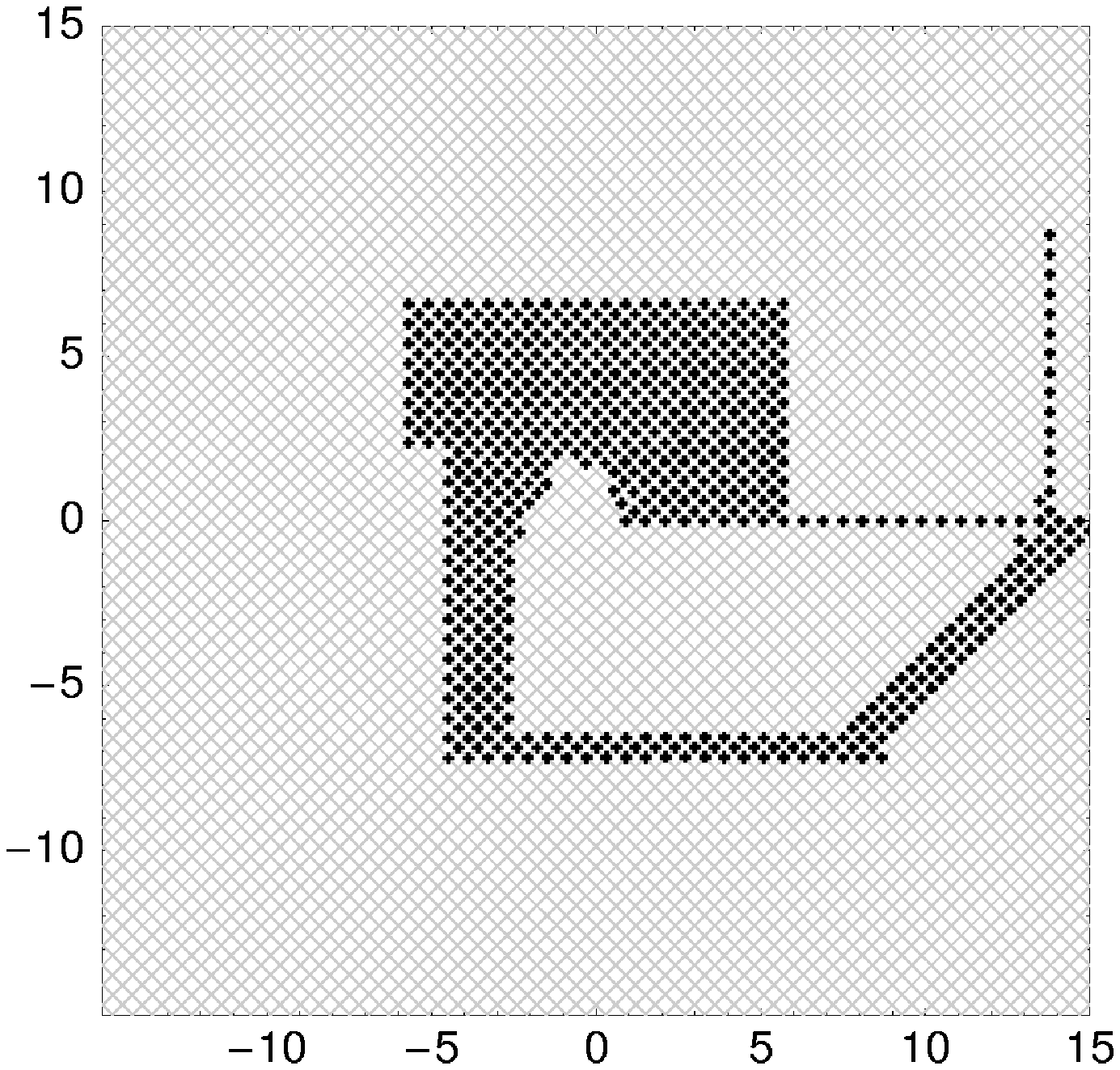}}
\put(77,2){\epsfysize5.5cm \epsfbox{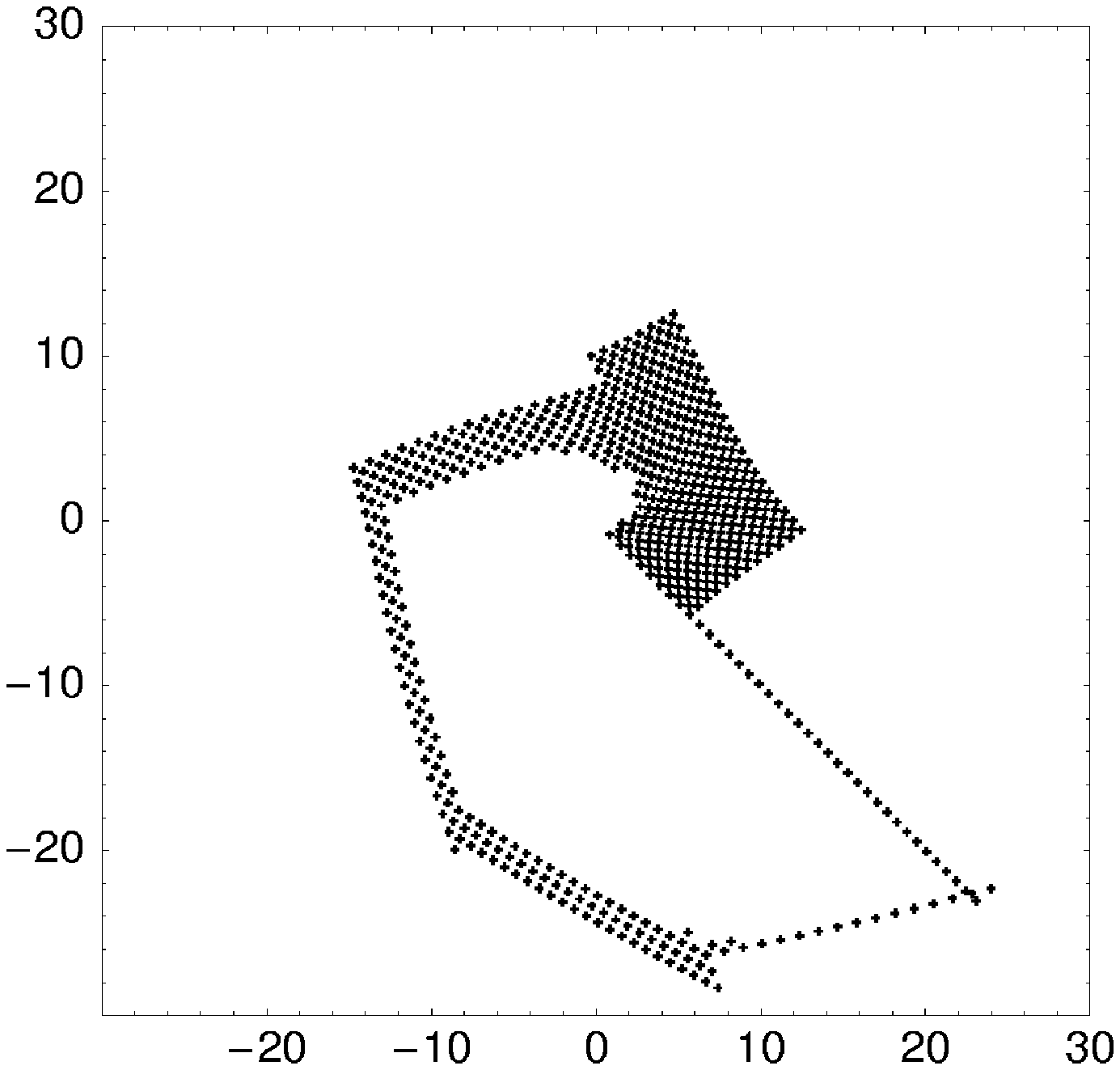}}
\put(35,-1){$\Re[q_4^{1/4}]$}
\put(9,27){\rotatebox{90}{$\Im[q_4^{1/4}]$}}
\put(100,-1){$\Re[q_3^{1/2}]$}
\put(74,27){\rotatebox{90}{$\Im[q_3^{1/2}]$}}
\end{picture}
}
\caption{Parts of the winding spectra of 
the conformal charges for $n_h=1$ with $\theta_4=0$. 
On the left panel the spectrum of $q_4^{1/4}$, while on the right
panel the spectrum of $q_3^{1/2}$. 
To show clearly winding correspondence we plot the WKB lattice with 
only some ``random'' vertices. However, the eigenvalues are related to all
vertices except vertices in the vicinity of $q_4=q_3=0$. 
 }
\label{fig:l43w01n1}
\end{figure}

\begin{table}[ht!]
\begin{center}
\begin{tabular}{|c|c|c|}
\hline
$q_3$ 	& $q_4$ 	& $E_4/4$ \\
\hline \hline
$	-1.36248i$&$ 0.67053	$&$	1.59563 $ \\ \hline
$	2.48686 + 0.45641i$&$ -0.673897 + 1.09776i	$&$	2.35448 $ \\ \hline
$	1.80352 + 1.62846i$&$ -0.599893 - 0.511892i	$&$	2.71937 $ \\ \hline
$	1.00697 + 2.71392i$&$ 1.24653 - 1.72595i	$&$2.97916 $ \\ \hline
$	-4.24348i$&$ 	4.86137$&$	-3.82033 $ \\ \hline
$	4.75145 + 0.852837i$&$-3.64269 + 2.82969 i	$&$	3.82631 $ \\ \hline
$	3.81467 + 2.64475i$&$ -3.76349 - 2.18733i	$&$	3.90119 $ \\ \hline
$	5.70668 - 1.4752i$&$ 0.745664 + 6.40784	i$&$	4.10895 $ \\ \hline
$	-2.79806 + 4.31936i$&$ 0.348195 + 6.24031i	$&$4.15134	 $ \\ \hline
$	6.35987 - 4.68879i$&$ 8.48213 + 6.07874 i	$&$4.61273	 $ \\ \hline
\end{tabular}
\end{center}
\caption{Winding lattice state spectrum for $n_h=1$ and $\theta_4=0$.}
\label{tab:l43w01n1}
\end{table}

Fig.~\ref{fig:l43s01n1}
with Table~\ref{tab:l43s01n1} show  an example of resemblant lattices 
where $\ell_3=0$
for $h=1$. We have an infinite number of such lattices and they can be enumerated
by an approximate ratio $|q_4^{1/4}/q_3^{1/2}|$
labelled by $\ell_4$.
On the other hand
in Fig.~\ref{fig:l43w01n1}
and Table~\ref{tab:l43w01n1} a part of a winding lattice for
$h=1$ is presented. 
These lattices are defined on a complicated winding Riemann surface
which has an infinite number of planes enumerated by $\ell_4$.
We have an infinite number of such lattices labelled
by $\ell_3\ne0$.

It turns out that for $n_h$ odd and $q_3=0$ 
trajectory-like solutions do not exist.
On the other hand, assuming that 
the ground state should not be degenerated \cite{Korchemsky:1994um},
the ground state should have $q_3=0$.
Therefore,
the ground state should belong to the $n_h$ even sector.
Indeed, the ground states for $N=4$ Reggeons has 
conformal Lorentz spin $n_h=0$ \cite{Korchemsky:2001nx}. 

\section{Point-like solutions}
\label{sec:psol}

It turns out that there are also 
other solutions for $\nu_h=0$
which do not lie on trajectories.
These specific solutions will be called in the following 
point-like solutions.
They appear for special values of conformal charges 
with $n_h \ne 0$
for
which 
in order to form
four independent solutions
less logarithmic terms are needed.
All these solutions do not have any continuation in the $\nu_h-$real space.

\subsection{Point-like solutions with even $n_h$
for $q_3=0$  and $\nu_h=0$}

For $h=(1+n_h)/2$ where $n_h \in 2 \mathbb{Z}$
there are two solutions with integer 
roots of the indicial equation (\ref{eq:b-exponents}), 
\ie $b_1=0$ and $b_2=1$,  
and two solutions with half-integer roots, 
\ie $b_3=4 s-h-1$ and $b_4=4 s+h-2$.
In this situation according to
(\ref{eq:v1q3s1b}) we have usually one $\Log(1-z)$ solution.
However, there are special values of conformal charges
for which recurrence relations produce the fourth independent
non-$\Log(1-z)$ solutions. 
In the case $n_h=2$ 
we have the following condition
\begin{equation}
64\,q_4-3+64\, q_3^2=0\,,
\label{eq:qq4q3n2}
\end{equation}
corresponding to the point-like solutions,
and for $n_h=4$
\begin{equation}
36864\, q_4^2 + (5760 + 40960\, q_3^2 )\,q_4 
-1215 + 1152\, q_3^2 + 4096\, q_3^4 =0\,.
\label{eq:qq4q3n4}
\end{equation}
Going farther, the conditions have a form
of polynomials of the order $n_h/2$  in $q_4$ and
$n_h$ in $q_3$.

\begin{figure}[ht]
\centerline{
\begin{picture}(100,50)
\put(5,0){\epsfysize5.3cm \epsfbox{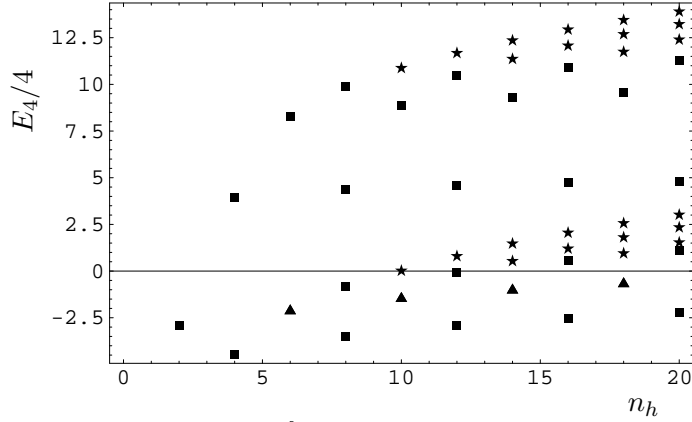}}
\put(82,-2){$n_h$}
\put(0,35){\rotatebox{90}{$E_4/4$}}
\end{picture}
}
\caption{The energy spectrum for 
point-like solutions with $n_h\in 2 \mathbb{Z}$ and $q_3=0$.
Here, shapes correspond to different degeneration of the spectrum.
}
\label{fig:Z2e}
\end{figure}

\begin{table}[ht!]
\begin{picture}(140,95)
\put(73,48){\scriptsize
\begin{tabular}{|c|c|c|}
\hline
$n_h$ & $q_4$ &  $E_4/4$ \\
\hline 
\hline
$8$&$4.793879   $&$-0.832250 $ \\
&$5.188999  $&$ 9.87791  $ \\
&$-0.777194$&$  -3.50838 $ \\
&$0.481816  $&$ 4.37493 $ \\ \hline
$12$&$-1.477392 $&$  -2.93906  $ \\
&$1.092692 $&$  4.59523 $ \\
&$7.441803  $&$ -0.05004  $ \\
&$8.672696 $&$  10.47907  $ \\
&$43.879856  $&$  0.80083  $ \\
&$43.921595   $&$ 11.67925  $ \\ \hline
$16$&$-2.359361 $&$  -2.53521  $ \\
&$1.973521 $&$  4.72661  $ \\
&$10.290397  $&$ 0.56333  $ \\
&$12.827684 $&$  10.91734  $ \\
&$62.689999  $&$ 1.20368 $ \\
&$62.835700 $&$  12.07543  $ \\
&$167.556849 $&$ 2.05481  $ \\
&$167.560211 $&$  12.93746  $ \\ 
\hline
$20$&$-3.412519$&$   -2.22192  $ \\
&$3.150012 $&$  4.80588 $ \\
&$13.265811 $&$  1.07723  $ \\
&$17.609886  $&$ 11.26143  $ \\
&$83.859313  $&$ 1.53510  $ \\
&$84.184263  $&$ 12.39907  $ \\
&$220.781025 $&$  2.34429   $ \\
&$220.794663  $&$ 13.22664  $ \\
&$451.493039 $&$  3.01760  $ \\
&$451.493257  $&$ 13.90039  $ \\
\hline
\end{tabular}}
\put(10,47){\scriptsize
\begin{tabular}{|c|c|c|}
\hline
$n_h$ & $q_4$ &  $E_4/4$ \\
\hline \hline
$\mybf{4}$&$\mybf{-0.275767}$&$\mybf{-4.48118}$ \\ 
&$0.119517 $&$  3.95157 $ \\ 
\hline
\hline
$n_h$ & $q_4$ &  $E_4/4$ \\
\hline \hline
$2$&$0.046875 	$&$	-2.93147$\\ \hline
$6$&$0.845724	$&$	8.27268$\\
&$0.147451 +0.545843 i $&$  -2.13355 $\\ 
&$0.147451 -0.545843 i $&$  -2.13355 $\\ \hline
$10$&$1.811503 		$&$	  8.88331$\\
&$17.565015  			$&$	 0.0193639 $\\
&$17.696871 			$&$	 10.87678 $\\
&$0.0804929 -1.267133 i $&$  -1.47238 $\\
&$0.0804929 +1.267133 i $&$  -1.47238 $\\ \hline
$14$&$3.085760  			$&$	 9.28477 $\\
&$26.493695			$&$	   0.537166 $\\
&$26.923075  			$&$	 11.36316 $\\
&$90.973271  			$&$	 1.47330 $\\
&$90.985515  			$&$	 12.35532 $\\
&$-0.066595 -2.190390 i $&$  	 -1.02327 $\\
&$-0.066595 +2.190390 i $&$	  -1.02327 $\\ \hline
$18$&$4.653448			$&$	   9.58289 $\\
&$36.621504			$&$	  0.95062 $\\
&$37.532130 			$&$	  11.74430$\\
&$124.243665  			$&$	 1.80846 $\\
&$124.289745 			$&$	12.68904 $\\
&$283.831799  		 	$&$	2.56459 $\\
&$283.832674 			$&$	13.44736 $\\
&$-0.291545 -3.302932 i $&$  -0.682788 $\\
&$-0.291545 +3.302932 i $&$  -0.682788 $\\
\hline  
\end{tabular}}
\end{picture}
\caption{The energy spectrum  and $q_4$ values for 
$n_h\in 2 \mathbb{Z}$ and  $q_3=0$.
The lowest energy solution is in bold.}
\label{tab:Z2q4}
\end{table}

Substituting $q_3=0$ to the above conditions
it turns out that in this case 
the quantization conditions (\ref{eq:C1-C0}) are satisfied.
Resulting solutions for $q_4$
as well as energies are given in 
Table~\ref{tab:Z2q4}.
All these solutions have $\theta_4=0$.  The energy minimum corresponds
to the  solution
with
\begin{equation}
h=\frac{5}{2},\quad q_3=0, \quad
q_4=-0.275767 \quad \mbox{with} \quad 
E_4/4=-4.448118\,.
\label{eq:groundstate}
\end{equation}
Note that solution (\ref{eq:groundstate}) has energy close
to two Pomerons: $2 E_2/4=-5.54518$
and it is lower than the Pomeron energy,
\ie with higher intercept.
For given $n_h \in 2\mathbb{Z}$ the
number of such solutions is $|n_h/2|$.
The energy spectrum is plotted in Fig.~\ref{fig:Z2e}.
One can see that increasing conformal Lorentz spin $n_h$ 
the energy goes up.

\subsection{Point-like solutions with even $n_h$ for $q_3 \ne 0$ and
$\nu_h=0$}

\begin{figure}[ht!]
\centerline{
\begin{picture}(140,57)
\put(12,2){\epsfysize5.5cm \epsfbox{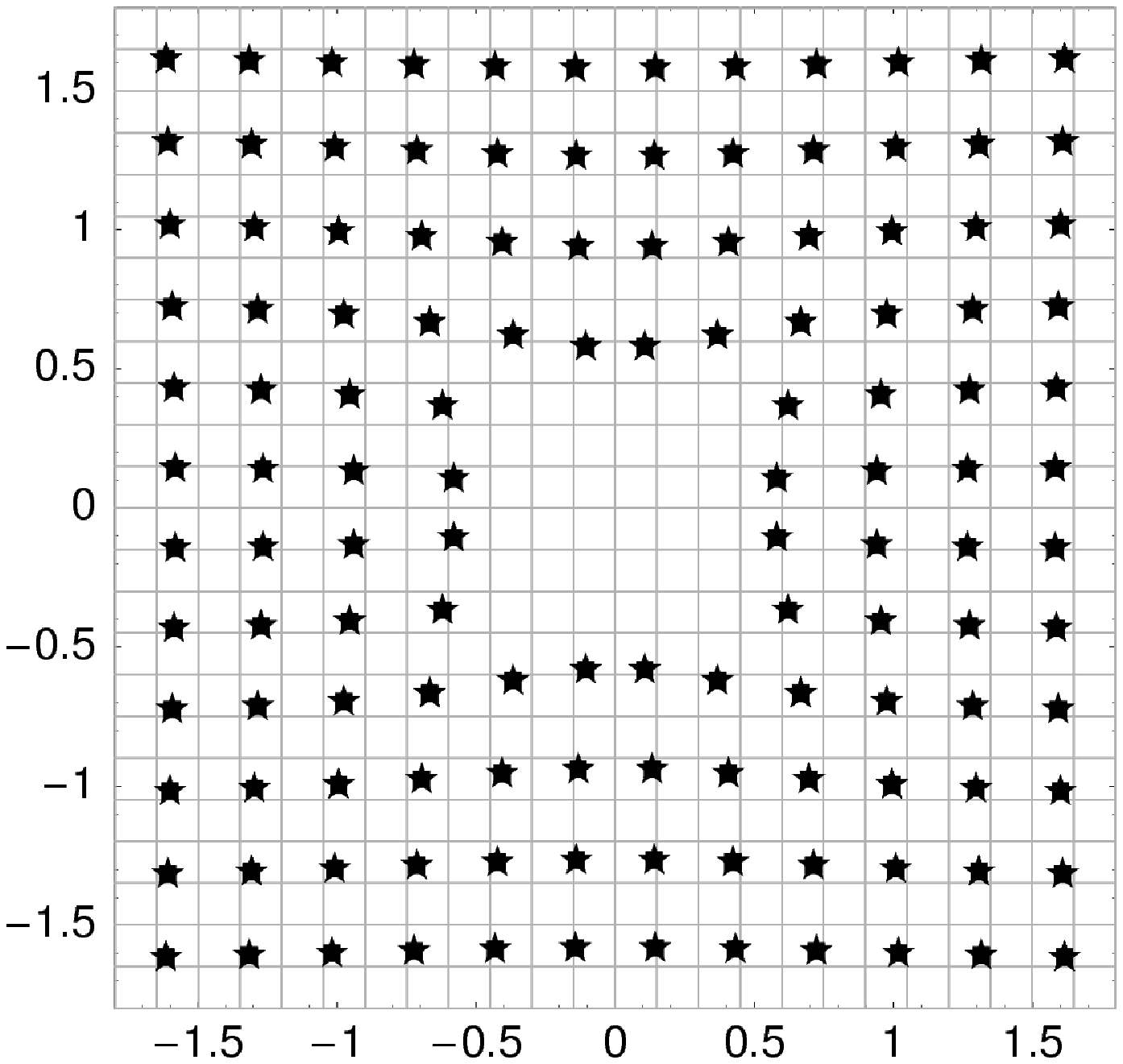}}
\put(72,2){\epsfysize5.5cm \epsfbox{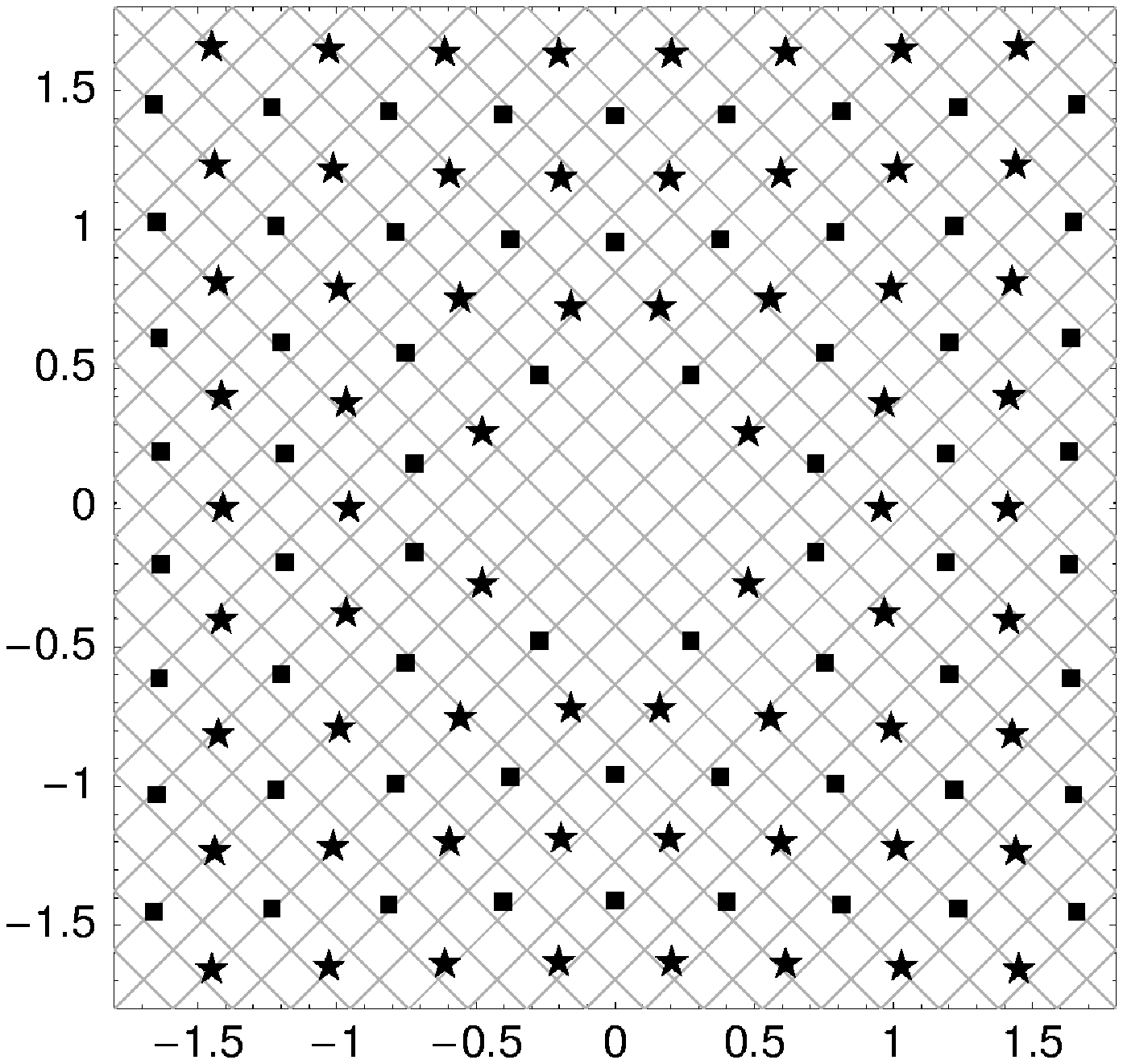}}
\put(35,-1){$\Re[q_4^{1/4}]$}
\put(8,25){\rotatebox{90}{$\Im[q_4^{1/4}]$}}
\put(95,-1){$\Re[q_3^{1/2}]$}
\put(68,25){\rotatebox{90}{$\Im[q_3^{1/2}]$}}
\end{picture}
}
\caption{The spectra of 
the conformal charges for  $n_h=2$ with 
$\theta_4=3 \pi/2$ ({\em stars}) and
$\theta_4= \pi/2$ ({\em boxes}).
On the left panel the spectrum of $q_4^{1/4}$, while on the right
panel the spectrum of $q_3^{1/2}$.}
\label{fig:l30p02}
\end{figure}

Additionally to solutions with $q_3=0$ we have also point-like solutions
with $q_3\ne0$.
Conformal charges of
these solutions can be plotted as vertices of one WKB lattice.
One can see their structure in Fig. \ref{fig:l30p02}.
The solutions with $n_h=2$ satisfy  conditions (\ref{eq:qq4q3n2}) and 
for large $q_4$ and $q_3$ WKB conditions
(\ref{eq:q4-quan}) and (\ref{eq:q3-quan}) are satisfied.
Condition (\ref{eq:qq4q3n2}) changes $q_4^{1/4}-$lattice to 
$q_3^{1/2}-$lattice 
rotated by $\pm \pi/4$ up to a small shift vanishing for large $q_k$.
Due to different roots in $q_3^{1/2}$ and $q_4^{1/4}$ as well as symmetry
$q_k \leftrightarrow q_k^{\star}$ two points in the $q_4^{1/4}-$lattice correspond
to one point in the $q_3^{1/2}-$lattice. On the other hand, points in the 
$q_4^{1/4}-$lattice are degenerated with respect
to $\theta_4= \pi/2, 3\pi/2$.
Conformal charges for a few first solutions are 
presented in Table \ref{tab:q3q4n2}. 

\begin{table}[ht!]
\begin{center}
\begin{tabular}{|c|c|c|}
\hline
$q_3$ 	& $q_4$ 	& $E_4/4$ \\
\hline \hline
$0.154083 -0.260344 i 	$&$0.0909123+0.0802291 i $&$ 0.0420484$ \\ \hline
$-0.494888+0.230748 i	$&$-0.144794+0.228388 i$&$ 1.05762$	\\ \hline
$-0.256542+0.839351 i	$&$	0.685571+0.430658$&$2.07086$ 	\\ \hline
$0.914503	$&$-0.78944	$&$ 2.10239$	\\ \hline
$0.791746+	0.730241 i$&$-0.0467357	-1.15633 i $&$ 2.45655$ \\ \hline
$-1.37191+0.462032 i	$&$-1.62178+	1.26773 i$&$ 3.04034$	\\ \hline
$0.357736+1.5676 i$&$	2.37626-1.12157 i$&$ 3.26625$	\\ \hline
$-1.0851+1.43035 i$&$ 0.91534+3.10414 i	$&$ 3.48118$	\\ \hline
$1.98735	$&$	-3.90267$&$3.67686$ 	\\ \hline
$1.83956+1.13559 i$&$	-2.04755-4.17796 i$&$3.84871$ 	\\ \hline

\end{tabular}
\end{center}
\caption{Point-like solution spectrum for $n_h=2$, $q_3 \ne0$ 
and $\theta_4=3\pi/2$.}
\label{tab:q3q4n2}
\end{table}
\begin{figure}[ht!]
\centerline{
\begin{picture}(140,57)
\put(12,2){\epsfysize5.5cm \epsfbox{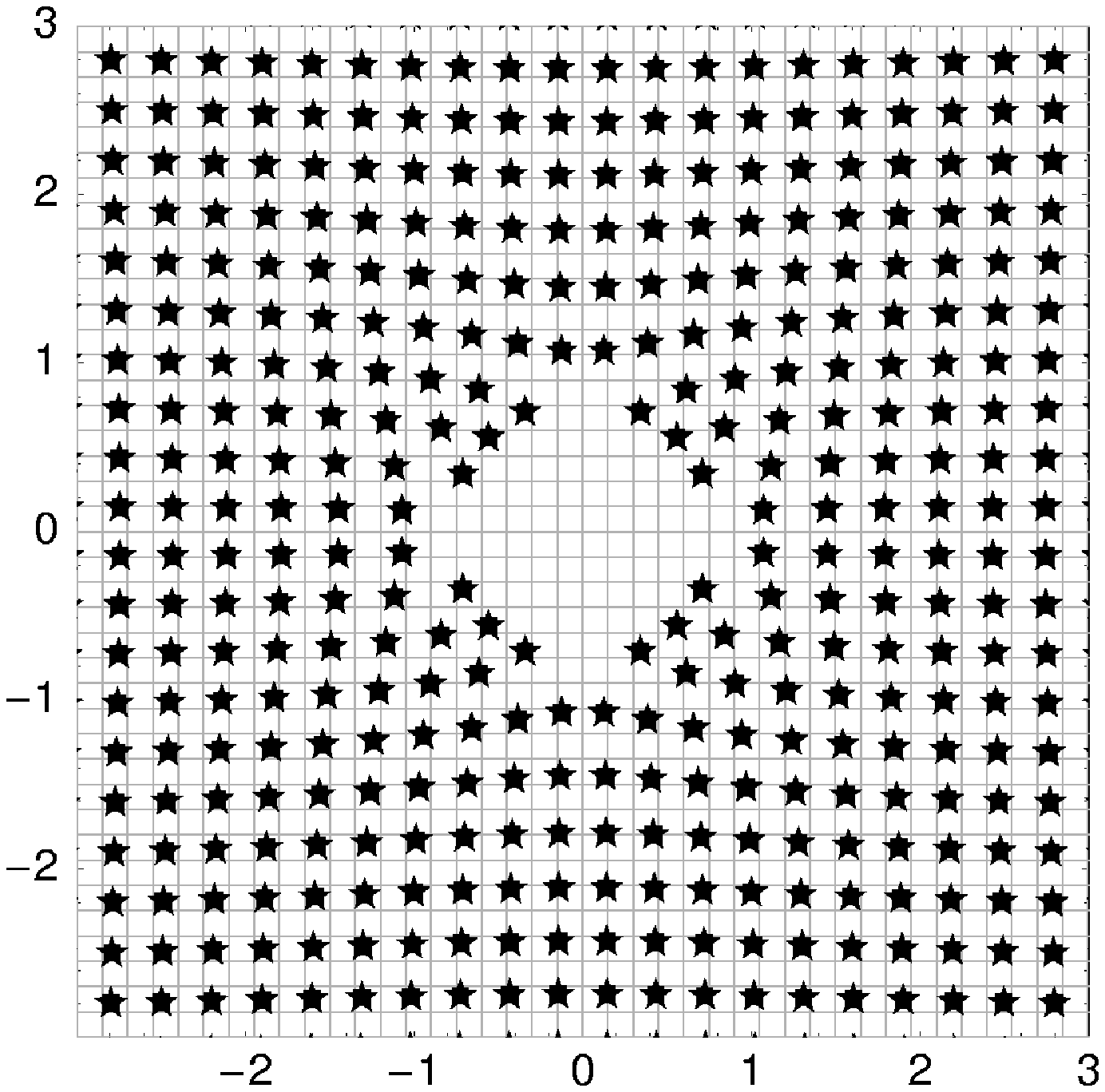}}
\put(72,2){\epsfysize5.5cm \epsfbox{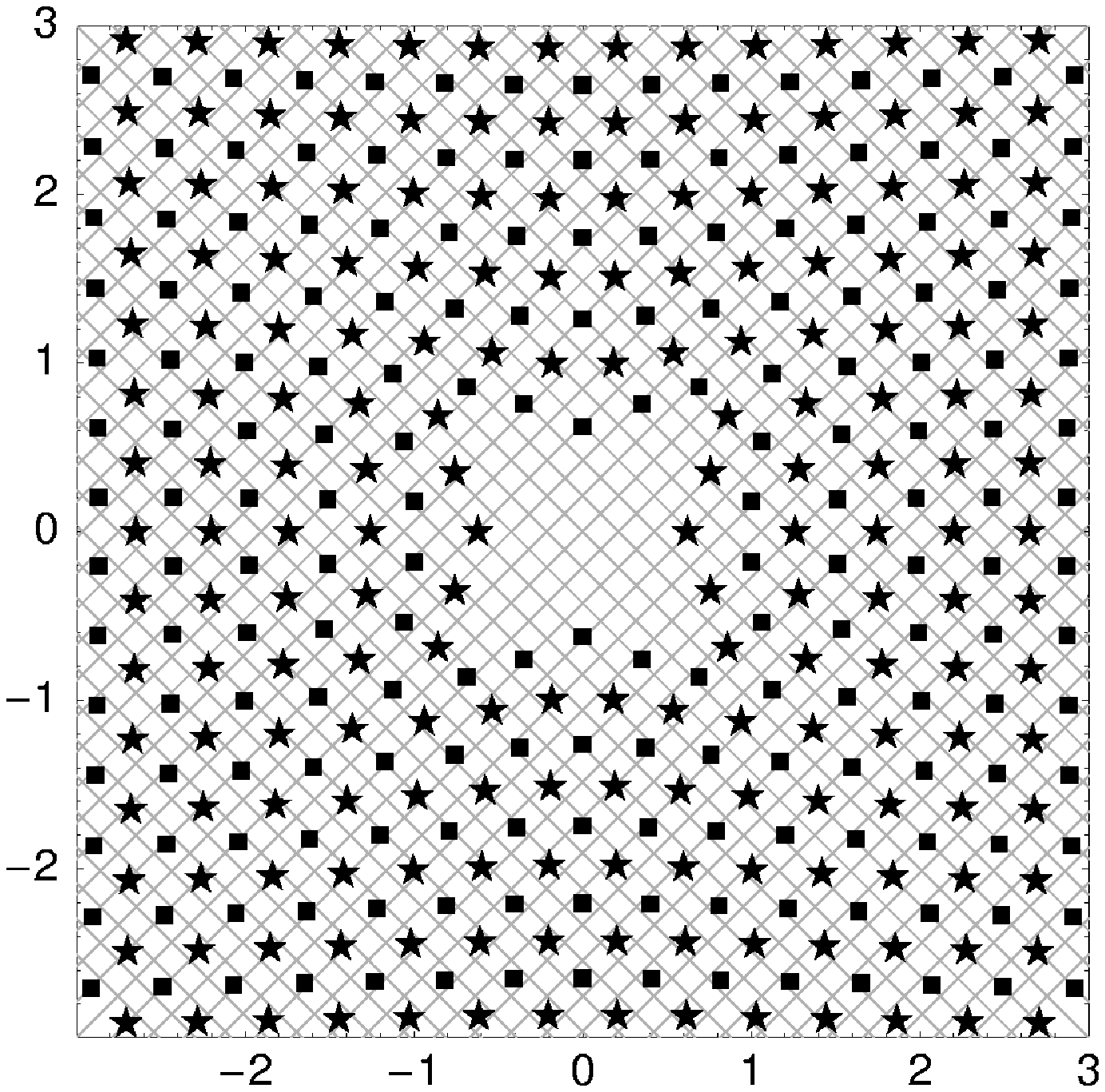}}
\put(35,-1){$\Re[q_4^{1/4}]$}
\put(8,25){\rotatebox{90}{$\Im[q_4^{1/4}]$}}
\put(95,-1){$\Re[q_3^{1/2}]$}
\put(68,25){\rotatebox{90}{$\Im[q_3^{1/2}]$}}
\end{picture}
}
\caption{The spectra of 
the conformal charges for $n_h=4$, $|\ell_3|=1$ with
$\theta_4=3\pi/2$ ({\em stars}) and
$\theta_4= \pi/2$ ({\em boxes}).
On the left panel the spectrum of $q_4^{1/4}$, while on the right
panel the spectrum of $q_3^{1/2}$.}
\label{fig:l30p11}
\end{figure}

\begin{table}[ht!]
\begin{center}
\begin{tabular}{|c|c|c|}
\hline
$q_3$ 	& $q_4$ 	& $E_4/4$ \\
\hline \hline
$ 	0.386612$&$	-0.388714 $&$ -0.393603   $ \\ \hline
$ 	0.450425 + 0.532463 i$&$	-0.0847377 - 0.377046 i $&$ 1.30653  $ \\ \hline
$ -0.961493 - 0.362284 i$&$ -0.961366 - 0.677412 i $&$ 2.13479   $ \\ \hline
$ 	0.264113 + 1.18142 i$&$	1.2114 - 0.611009 i $&$  2.91603  $ \\ \hline
$ 	-0.83545 - 1.13937 i$&$	0.46445 - 1.88396 i $&$ 3.0377   $ \\ \hline
$ 	1.5859$&$	-2.67051 $&$  3.10799  $ \\ \hline
$ 1.50126 + 0.951915 i	$&$	-1.49397 - 2.84724 i $&$ 3.4103   $ \\ \hline

\end{tabular}
\end{center}
\caption{Point-like solution spectrum for $n_h=4$, $|\ell_3|=1$, 
$q_3 \ne0$ and $\theta_4=3\pi/2$.}
\label{tab:q3q4n4l1}
\end{table}
\begin{figure}[ht!]
\centerline{
\begin{picture}(140,57)
\put(12,2){\epsfysize5.5cm \epsfbox{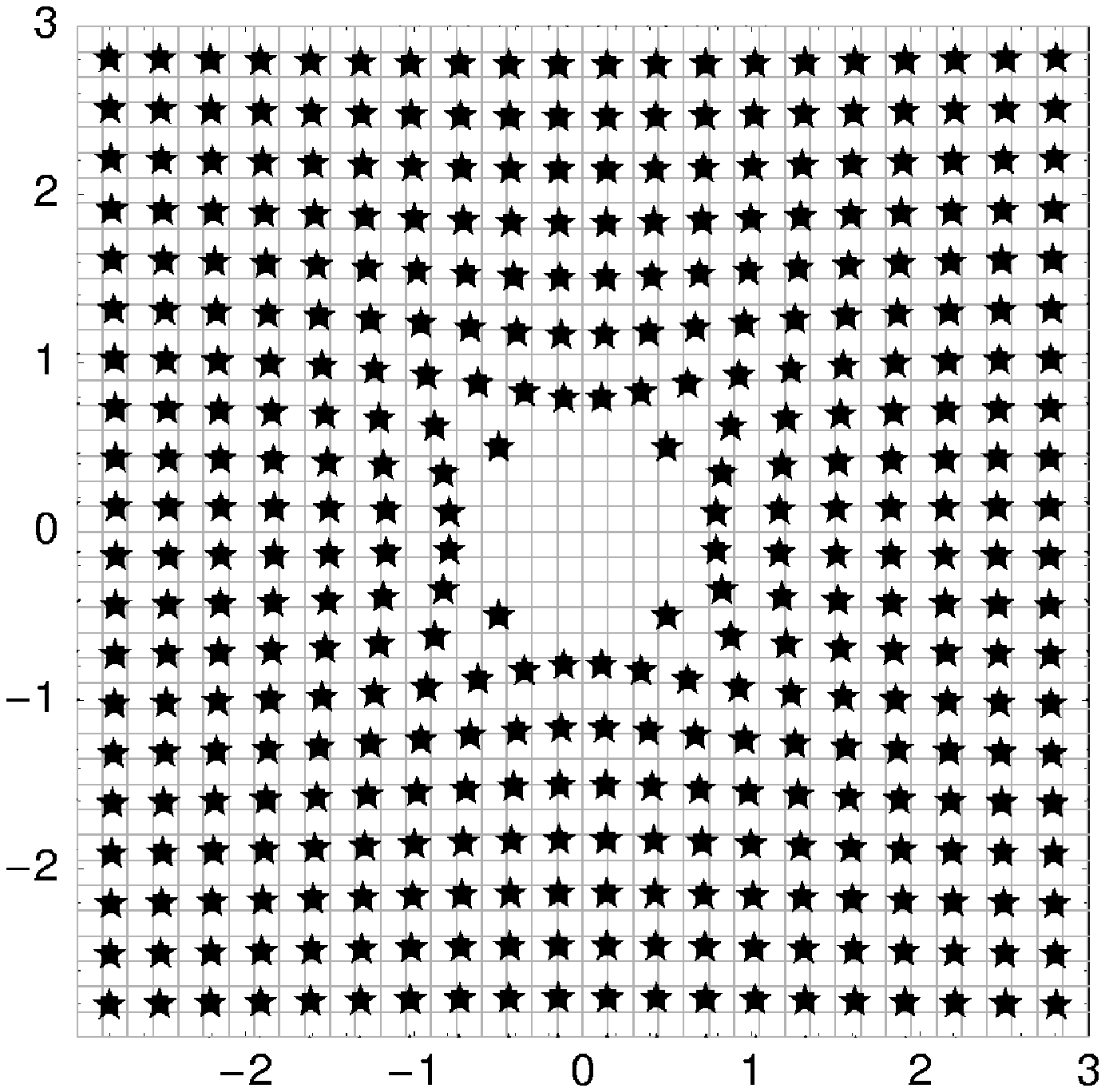}}
\put(72,2){\epsfysize5.5cm \epsfbox{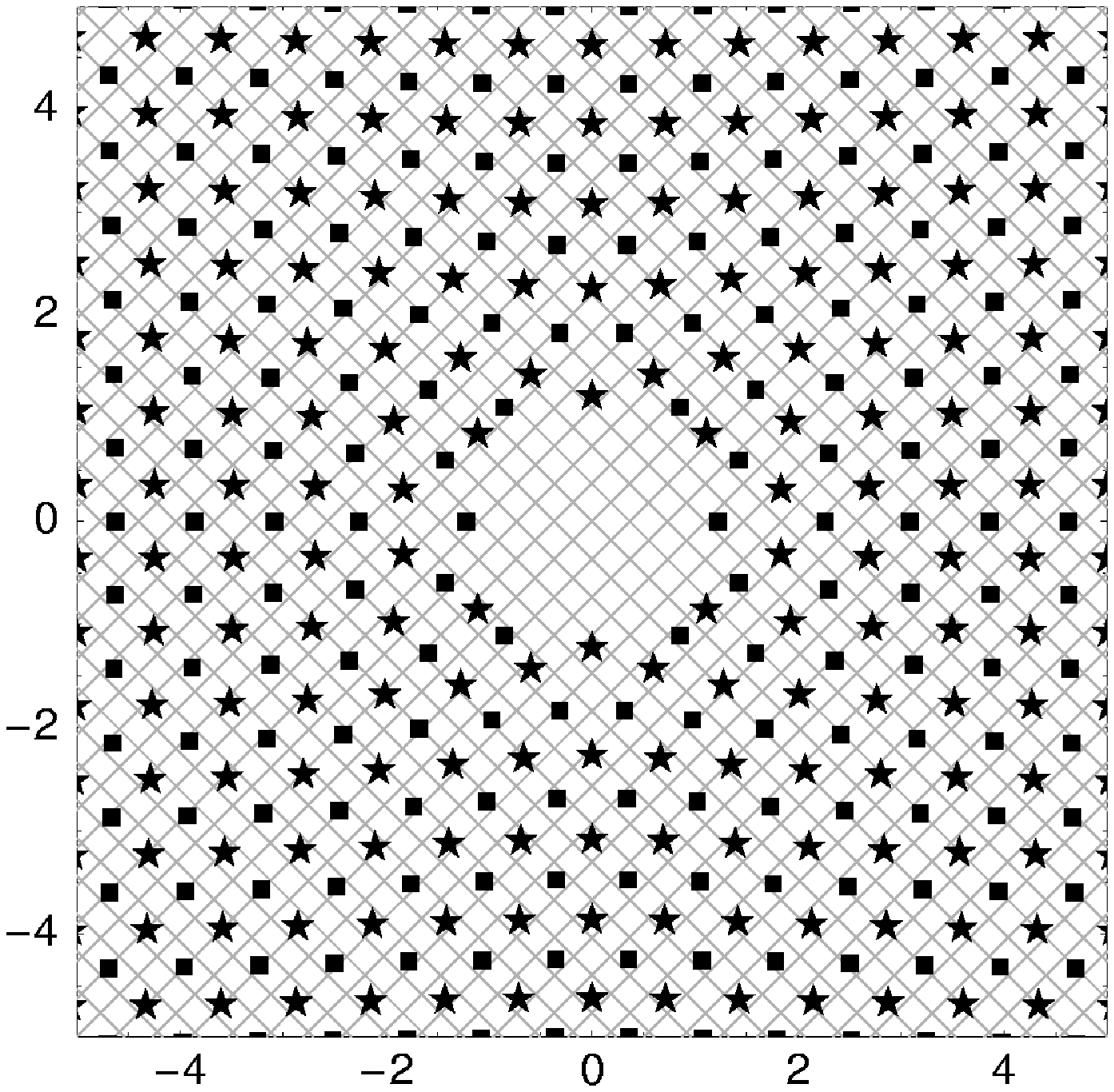}}
\put(35,-1){$\Re[q_4^{1/4}]$}
\put(8,25){\rotatebox{90}{$\Im[q_4^{1/4}]$}}
\put(95,-1){$\Re[q_3^{1/2}]$}
\put(68,25){\rotatebox{90}{$\Im[q_3^{1/2}]$}}
\end{picture}
}
\caption{The spectra of 
the conformal charges for  $n_h=4$, $|\ell_3|=3$ with
$\theta_4=3 \pi/2$ ({\em stars}) and
$\theta_4= \pi/2$ ({\em boxes}).
On the left panel, where {\em boxes} coincide with {\em stars}, 
the spectrum of $q_4^{1/4}$, while on the right
panel the spectrum of $q_3^{1/2}$}
\label{fig:l30p13}
\end{figure}
\begin{table}[ht!]
\begin{center}
\begin{tabular}{|c|c|c|}
\hline
$q_3$ 	& $q_4$ 	& $E_4/4$ \\
\hline \hline
$0.502865 + 1.89451 i$&$	0.345943 - 0.217117 i $&$ 1.05203  $ \\ \hline
$-1.49327$&$	-0.246831 $&$ 1.80272   $ \\ \hline
$-1.68063 - 1.69926 i 	$&$-0.0085373 - 0.641561 i $&$1.83642    $ \\ \hline
$ 3.26053 - 1.1522 i	$&$-1.04678 + 0.836881 i $&$ 2.71529   $ \\ \hline
$ -0.890225 - 4.04854 i$&$1.7154 - 0.801894 i $&$ 2.85804  $ \\ \hline
$ 2.75764 + 3.74489 i	$&$	0.697135 - 2.29665 i $&$ 3.15922   $ \\ \hline
$-5.12753$&$	-2.93541 $&$ 3.45247   $ \\ \hline
\end{tabular}
\end{center}
\caption{Point-like solution spectrum for $n_h=4$, $|\ell_3|=1$, 
$q_3 \ne0$ and $\theta_4=3 \pi/2$.}
\label{tab:q3q4n4l3}
\end{table}
For $n_h=4$ we have point-like solutions whose conformal charges 
satisfy (\ref{eq:qq4q3n4}). Taking this condition for $q_4 \to \infty$
and ${q_3}^2/q_4 \sim 1$
we obtain
\begin{equation}
9 q_4^2+10 q_3^2 q_4+q_3^4=0
\end{equation}
which gives two solutions.
The first solution, $q_3^2=-9 q_4$, is related to a lattice with
$|\ell_3|=3$ while the second one,  $q_3^2=- q_4$, corresponds to
a lattice with $|\ell_3|=1$.
The lattice with $|\ell_3|=1$ is plotted in Fig. \ref{fig:l30p11} and Table
\ref{tab:q3q4n4l1}
and the lattice with $|\ell_3|=3$ in Fig. \ref{fig:l30p13} and Table
\ref{tab:q3q4n4l3}.

Moreover, additionally to these point-like solutions 
one can find
descendent solutions, which have $q_4=0$, \eg a point-like solution 
with $h=5/2$ where 
\begin{equation}
q_3=\pm i 3 \sqrt{5}/8, 
\quad \theta_4=\pi,
\quad E_4/4=-1.158883\,,
\label{eq:nhdesc}
\end{equation}
or with $h=7/2$ where
\begin{equation}
q_3=\pm  3 \sqrt{35}/8, 
\quad \theta_4=\pi,
\quad E_4/4=5.50778\,.
\label{eq:nhdesc2}
\end{equation}
The descendent point-like solutions for $N=4$
reggeized gluons come
from the $N=3$ point-like solutions with the 
same quantum numbers and 
the same energy\footnote{If the $N=3$ point-like solutions 
with the quantum numbers (\ref{eq:nhdesc}) had been
normalizable the odderon intercept of 
the three reggeized gluon
would be higher than one.}. 

\subsection{Point-like solutions with odd $n_h$ for $\nu_h=0$}

\begin{figure}[ht]
\centerline{
\begin{picture}(140,57)
\put(12,2){\epsfysize5.5cm \epsfbox{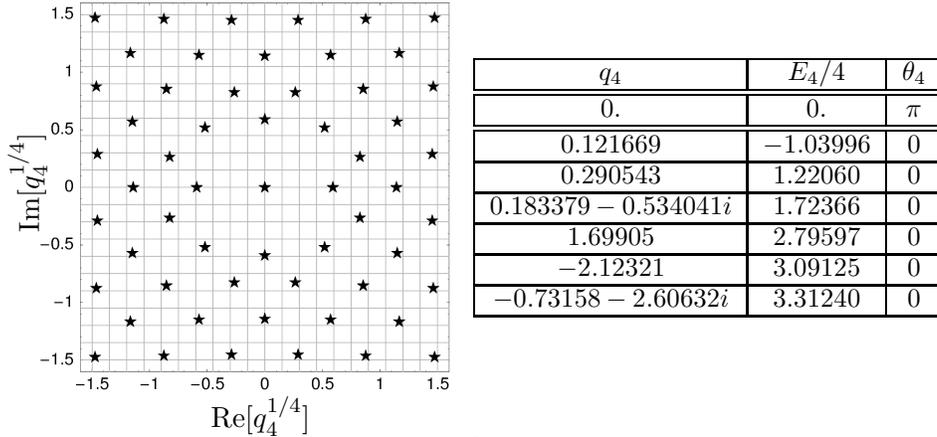}}
\put(35,-1){$\Re[q_4^{1/4}]$}
\put(8,25){\rotatebox{90}{$\Im[q_4^{1/4}]$}}
\put(70,30){\small
\begin{tabular}{|c|c|c|}
\hline
$q_4$ & 
${E_4/4}$ & $\theta_4$ \\
\hline
\hline
$0. $ &  $0.$ & $\pi$\\ \hline \hline
$ 0.121669$ &  $-1.03996$ &$0$\\ \hline
$0.290543 $ & $1.22060$ &$0$ \\ \hline
$0.183379- 0.534041i $ &  $1.72366$ &$0$\\ \hline 
$1.69905 $  &   $2.79597$ &$0$\\ \hline
$-2.12321 $ & $3.09125$ &$0$ \\ \hline
$-0.73158- 2.60632i$ & $3.31240$&$0$ \\ \hline
\end{tabular}
}
\end{picture}
}
\caption{The spectrum of quantized $q_4^{1/4}$ for the 
system  with $n_h=1$ and $q_3=0$.}
\label{fig:pln1}
\end{figure}

For $n_h=1$ and $q_3=0$
there is one lattice for point-like solutions with:
\begin{eqnarray}
&&v_m(y) = y^{b_m} \sum_{n=0}^{\infty} v_n^{(m)} y^n\,,
\nonumber
\\[2mm]
&&v_{m+2}(y) = y^{b_{m+2}} \sum_{n=0}^{\infty} v_n^{(m+2)} y_n+ \Log(y) v_m(y)\,,
\label{eq:v1q3s1c}
\end{eqnarray}
where  $m=1,2$, while $b_1=1$, $b_2=0$, $b_3=-1$, $b_4=-2$ for $s=0$ 
and $b_1=3$, $b_2=2$, $b_3=1$, $b_4=0$ for $s=1$.
The corresponding quantum values of $q_4$ and $E_4/4$ for point-like 
solutions are 
plotted in Fig.~\ref{fig:pln1}.
One can see that the solution with the lowest energy $E_4/4=-1.03996$ 
corresponds to  
$(\ell_1,\ell_2)=(2,0)$.
Additionally, we have solutions with $q_2=q_3=q_4=0$
and $E_4=0$ as well as $\theta_4=\pi$ 
with the $z=1$ solutions 
$Q_i^{(1)}(z)=z^{1-s} (1-z)^{b_i} \sum_{n=0}^{\infty} v_n^{(i)} (1-z)^n$.

\begin{figure}[ht]
\centerline{
\begin{picture}(140,57)
\put(12,2){\epsfysize5.5cm \epsfbox{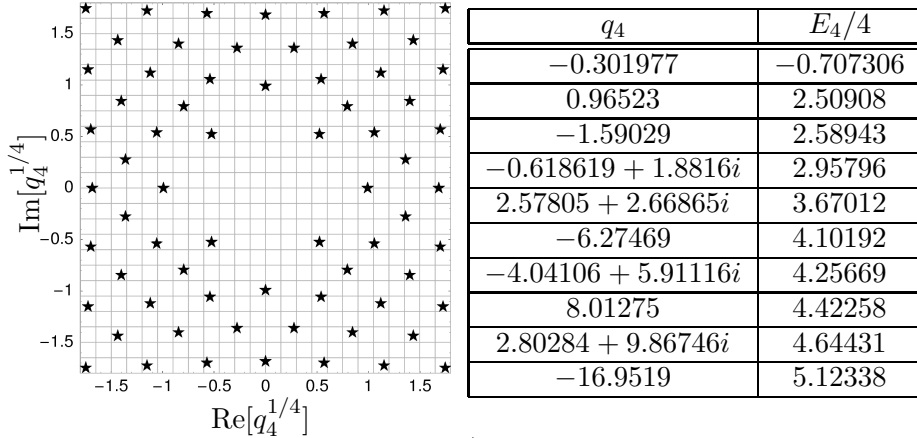}}
\put(35,-1){$\Re[q_4^{1/4}]$}
\put(8,25){\rotatebox{90}{$\Im[q_4^{1/4}]$}}
\put(68,28){
\begin{tabular}{|c|c|}
\hline
$q_4$ 	& $E_4/4$ \\
\hline \hline
$-0.301977 $&$-0.707306 $ \\ \hline
$0.96523 $&$2.50908 $ \\ \hline
$-1.59029 $&$2.58943 $ \\ \hline
$-0.618619 + 1.8816 i  $&$2.95796 $ \\ \hline
$2.57805 + 2.66865 i $&$3.67012 $ \\ \hline
$-6.27469 $&$4.10192 $ \\ \hline
$-4.04106 + 5.91116i $&$4.25669 $ \\ \hline
$8.01275 $&$4.42258 $ \\ \hline
$2.80284 + 9.86746 i $&$4.64431 $ \\ \hline
$-16.9519 $&$5.12338 $ \\ \hline
\end{tabular}
}
\end{picture}
}
\caption{The spectrum of quantized $q_4^{1/4}$ for 
point-like solutions with $n_h=3$ and $q_3=0$.}
\label{fig:pln0}
\end{figure}

\begin{figure}[ht!]
\centerline{
\begin{picture}(140,57)
\put(12,2){\epsfysize5.5cm \epsfbox{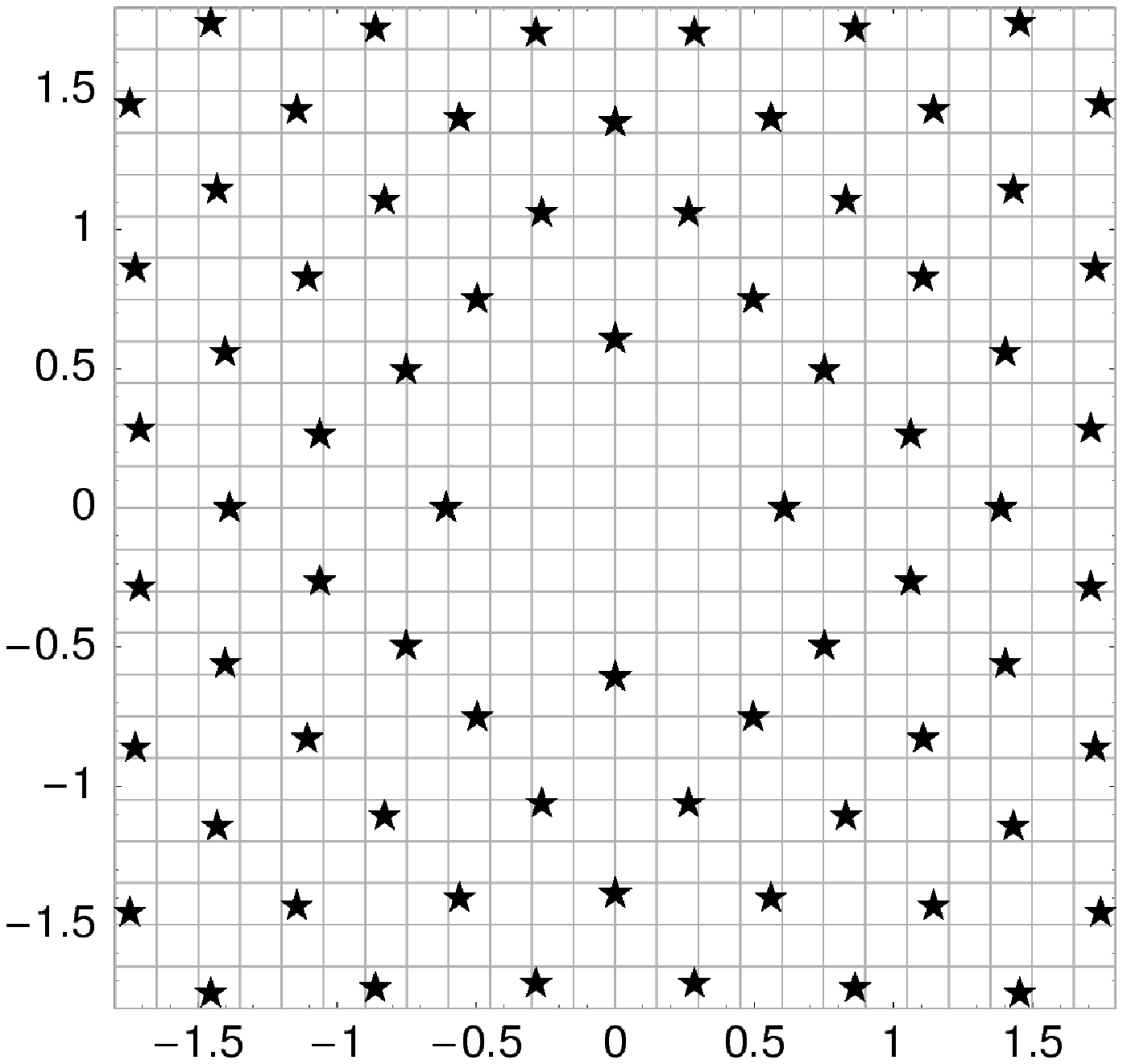}}
\put(72,2){\epsfysize5.5cm \epsfbox{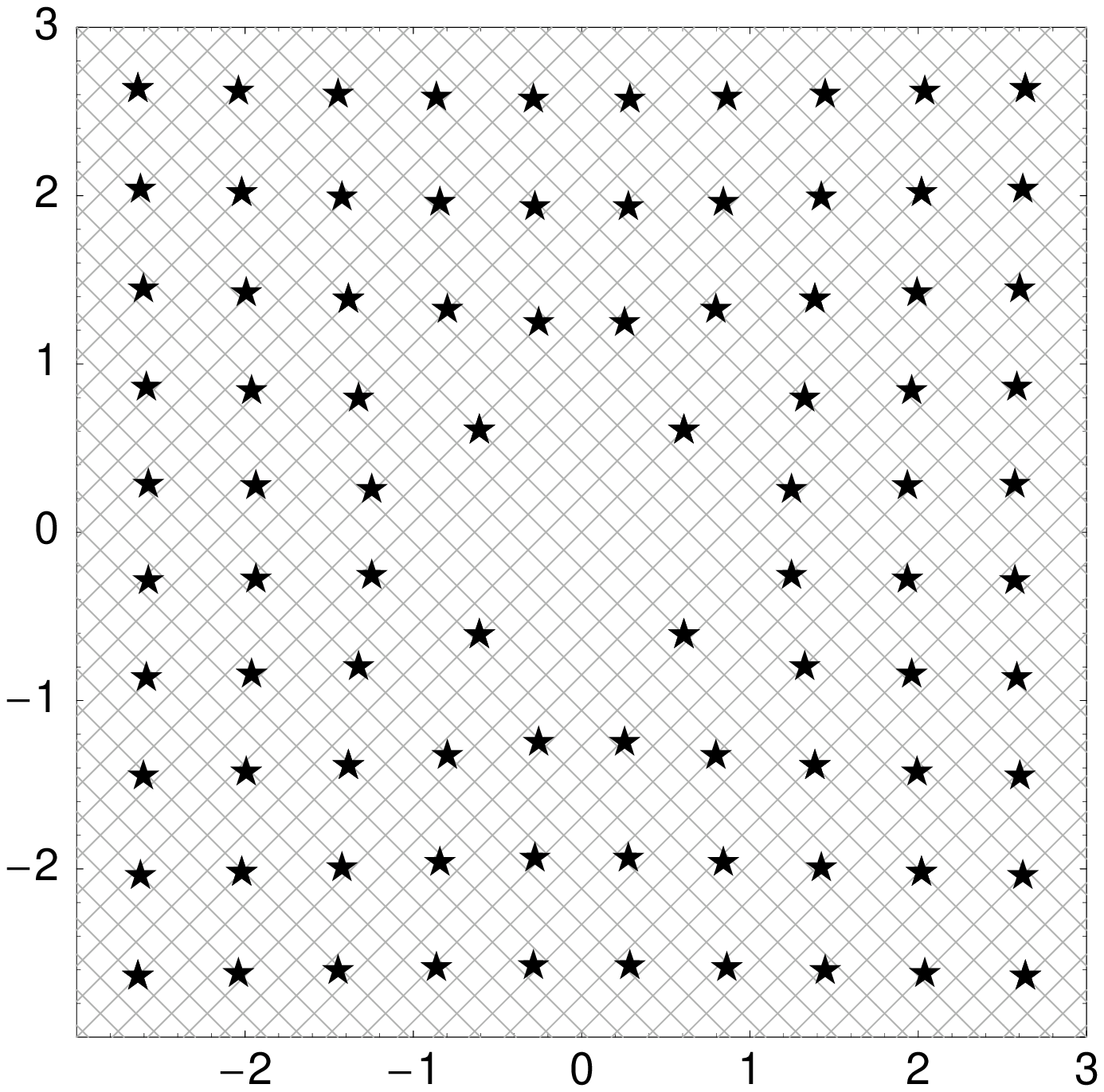}}
\put(35,-1){$\Re[q_4^{1/4}]$}
\put(8,25){\rotatebox{90}{$\Im[q_4^{1/4}]$}}
\put(95,-1){$\Re[q_3^{1/2}]$}
\put(68,25){\rotatebox{90}{$\Im[q_3^{1/2}]$}}
\end{picture}
}
\caption{Point-like rotated spectra of 
the conformal charges for $n_h=3$ with $\theta_4=0$. 
On the left panel the spectrum of $q_4^{1/4}$, while on the right
panel the spectrum of $q_3^{1/2}$}
\label{fig:l43r01n3}
\end{figure}

For odd $|n_h| >1$ we have 
also WKB lattices of point-like solutions.
For example 
point-like solutions with $q_3=0$ are plotted in Fig.~\ref{fig:pln0}.
On the other hand, Fig.~\ref{fig:l43r01n3}
and Table~\ref{tab:l43r01n3} show an example of point-like rotated  
lattices. 

\begin{table}[ht!]
\begin{center}
\begin{tabular}{|c|c|c|}
\hline
$q_3$ 	& $q_4$ 	& $E_4/4$ \\
\hline \hline
$   -0.73865  i  $&$	0.136401$&$	0.634391$ \\ \hline
$   1.48983 - 0.636904 i   $&$-0.453486 + 0.474439i 	$&$ 1.96712$ \\ \hline
$   1.12055 - 2.11477i    $&$	0.804151 + 1.18485i $&$2.63622	$ \\ \hline
$  3.84094 i    $&$	3.68821$&$	3.59751$ \\ \hline
$   3.66894 - 1.07472  i  $&$	-3.07653 + 1.97155i$&$	3.62067$ \\ \hline
$     3.13679 - 3.30259 i $&$	0.266913 + 5.17977i$&$	3.95358$ \\ \hline
$      1.94306 - 5.67594i $&$	7.11021 + 5.51436i$&$	4.50006$ \\ \hline
$     6.54514 - 1.4747 i $&$	-10.166 + 4.82605i$&$	4.73498$ \\ \hline
$     5.94333 - 4.46239 i  $&$-3.85257 + 13.2607i	$&$4.93545$ \\ \hline
$     8.1565i  $&$	16.6321$&$	5.11626$ \\ \hline
\end{tabular}
\end{center}
\caption{Point-like solution spectrum for  $n_h=3$ and $\theta_4=0$.}
\label{tab:l43r01n3}
\end{table}

\section{Examples of norm calculation}

\begin{figure}[ht!]
\centerline{
\begin{picture}(140,45)
\put(11,2){\epsfysize4.5cm \epsfbox{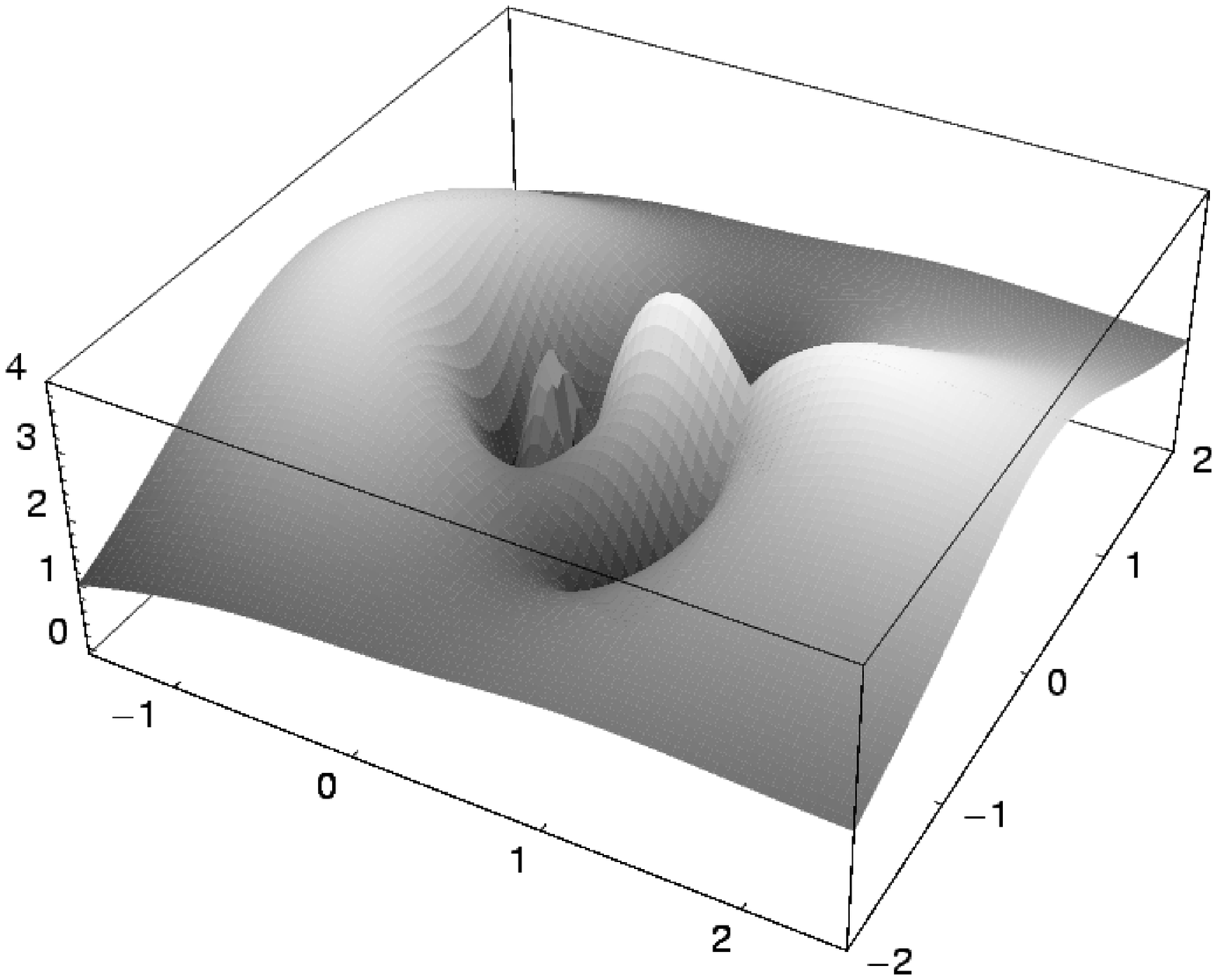}}
\put(72,2){\epsfysize4.5cm \epsfbox{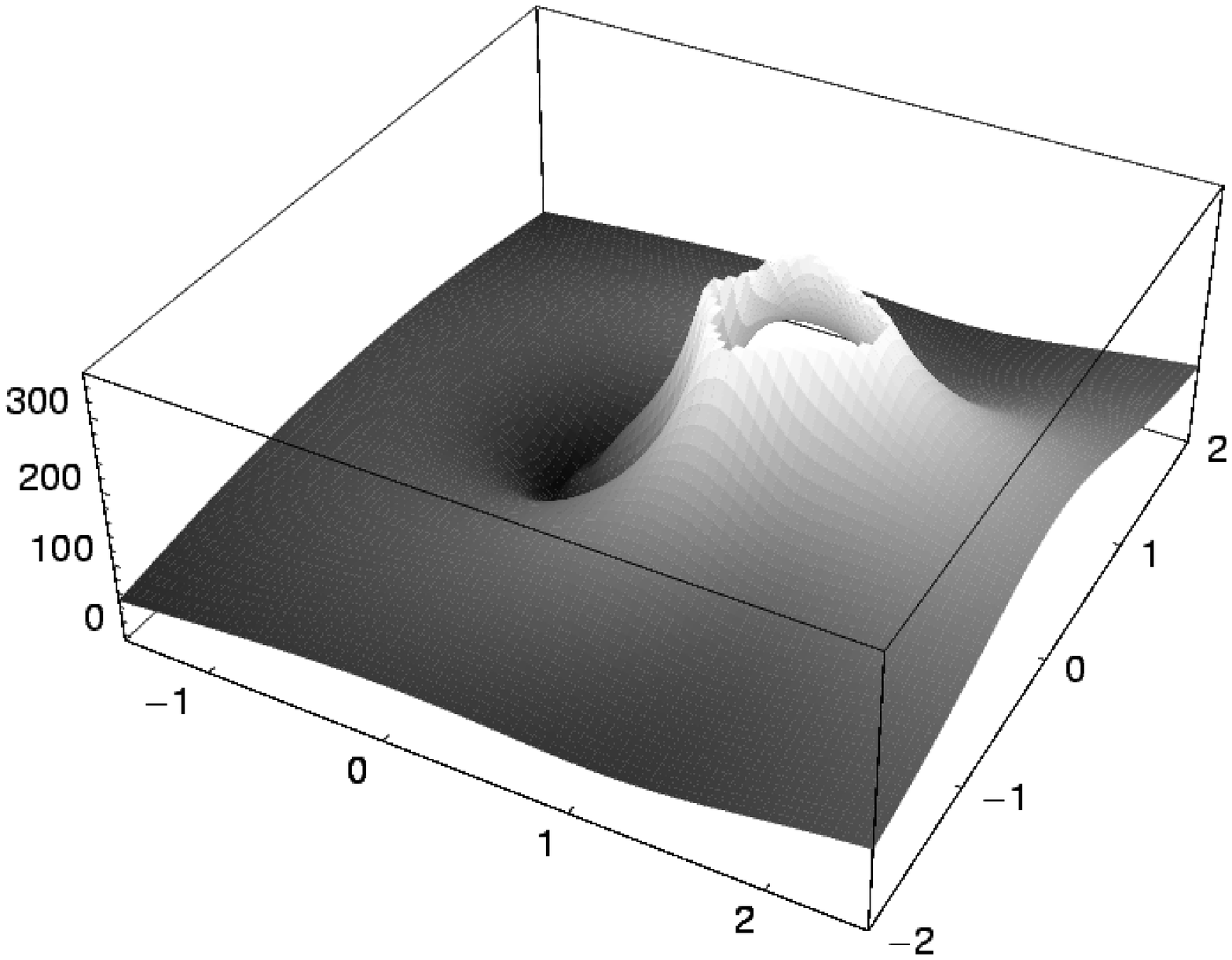}}
\put(28,5){$x$}
\put(60,12){$y$}
\put(6,17){\rotatebox{90}{$|\tilde Q_{q,\wbar q}(z,z^{\ast})|$}}
\put(93,5){$x$}
\put(120,12){$y$}
\put(67,20){\rotatebox{90}{$|\tilde Q_{q,\wbar q}(z,z^{\ast})|$}}
\end{picture}
}
\caption{The Baxter eigenfunctions $\tilde Q_{q,\wbar q}(z,z^{\ast})$ with 
$z=x+iy$.The trajectory-like solution with 
$h=3/2$, $q_3=0$ and $q_4=1.4100348$ is plotted
on the left panel, while on the right panel
we plot the point-like solution with $h=3/2$, $q_3=0$ and $q_4=3/64$.}
\label{fig:phi}
\end{figure}

The eigenfunctions $\tilde Q_{q,\wbar q}(z,z^{\ast})$
presented in Section 5 whose conformal charges form 
trajectories
are finite on the whole complex $z-$plane.
One of such functions is shown on the left panel in Fig.~{\ref{fig:phi}}.
The norm of these trajectory solutions is infinite for $\nu_h=\nu_h'$.
However, they describe continuous spectra so it is natural that they 
are normalized to the Dirac delta function (\ref{eq:normqq}).

On the other
hand there 
exist solutions that
do not form trajectories in real $\nu_{h}$.
An example of the point-like eigenfunction $\tilde Q_{q,\wbar q}(z,z^{\ast})$
with $h=3/2$, $q_3=0$ and $q_4=3/64$ 
is plotted on the right panel
in Fig.~{\ref{fig:phi}}.
One can see a strong singularity at $z=z^{\ast}=1$,
which is present for all point-like solutions.
These solutions
appear only for $\nu_{h}=0$, therefore, they have to be normalizable to
the Kronecker delta symbol while their norm (\ref{eq:normzz})
has to be finite for all quantum
numbers. 
It turns out that only series solutions around $z=1$ can give 
infinite contribution to the norm (\ref{eq:normzz}).
Series around $z=0$ and $z=\infty$ due to their asymptotics give at most 
finite contribution.
Let us therefore 
consider calculation of contribution to norm from solutions around 
$z=1$.

\subsection{Point-like solutions with odd $n_{h}$}

The solution for $h=1$ and $q_{3}=0$ and $q_{4}\ne0$ around $z=1$ has a form
$\tilde{Q}_{q,\wbar q}(z,\wbar{z})=\wbar{Q}_{i}(\wbar{z})C^{(1)}_{ij}Q_{j}(z)$,
with $Q_{i}(z)=z^{(1-s)}v_{i}(1-z)$
and (\ref{eq:v1q3s1c}).
From singlevaluedness condition we get that: 
\begin{equation}
C^{(1)}=\left(
\begin{array}{cccc}
X_1 & X_2 & X_5 & X_7\\
X_3 & X_4 & X_6 & X_8\\
X_5 & X_7 & 0 & 0\\
X_6 & X_8 & 0 & 0
\end{array}\right)\:.
\end{equation}
From numerical calculation one can see that $X_6=X_7=0$ and $X_2=X_3$. The
other ones may be different from zero. The most singular contribution
comes from the term
\begin{equation}
\tilde{Q}_{q,\wbar q}(z,\wbar{z})=X_8\: z\: 
v_{4}(1-z)\wbar{v}_{2}(1-\wbar{z})+\ldots=X_8'\: 
z(1-z)^{-2}(1-\wbar{z})^{2}+\ldots\:,
\label{eq:Qzzsin}
\end{equation}
where $X_8'=X_8 h_{0}^{(4)}\wbar{h}_{0}^{(2)}$. Substituting (\ref{eq:Qzzsin})
to (\ref{eq:normzz}) one obtains
\begin{multline}
\langle q,\wbar q|q,\wbar q\rangle=48\: D^{6}(\prod_{i=1}^{3}\int_{\rho}^{R}\rho_{i}d\rho_{i}\int_{0}^{2\pi}d\phi_{i})\\
\times
\frac{\prod_{j<k}(\rho_{j}\textrm{e}^{i\phi_{j}}-\rho_{k}\textrm{e}^{i\phi_{k}})(\rho_{j}\textrm{e}^{-i\phi_{j}}-\rho_{k}\textrm{e}^{-i\phi_{k}})}{\rho_{1}^{4}\rho_{2}^{4}\rho_{3}^{4}}+\ldots\:,
\end{multline}
with $D\sim X_8'$ and where we perform the integrals around $z_{i}=\wbar{z}_{i}=1$
from radius $\rho\to0$ up to radius $R$ with $z_{j}=1+\rho_{j}\textrm{e}^{i\phi_{j}}$
and $\wbar{z}_{i}=1+\rho_{j}\textrm{e}^{-i\phi_{j}}$ . Evaluating the 
integrals one gets
\begin{equation}
\langle q,\wbar q|q,\wbar q\rangle\sim\lim_{\rho\to0}R^{2}\frac{\log(\rho/R)}{\rho^{2}}+\ldots\log(\rho/R)+\ldots\,,
\end{equation}
\ie the singularity at $\rho\to0$. The other terms give finite results 
around $\rho=0$ so the above singularity cannot be cancelled.
Such solutions are non-normalizable. Similarly one can calculate
the norm of the other point-like solutions with $n_{h}\in2\mathbb{Z}+1$.

\subsection{Point-like solutions with even $n_{h}$}

For point-like solutions with even $n_{h}$ we can
perform similar calculations 
but here
we don't have logarithmic solutions. The solution around $z=1$ has
a form:
$\tilde{Q}_{q,\wbar q}(z,\wbar{z})=\wbar{Q}_{i}(\wbar{z})C^{(1)}_{ij}Q_{j}(z)$,
with $Q_{i}(z)=z^{(1-s)}v_{i}(1-z)$ ,
where
\begin{equation}
v_{i}(y)=y^{b_{i}}\sum_{n=0}h_{n}^{(i)}y^{n}\,,
\end{equation}
with $b_{1}>b_{2}>b_{3}>b_{4}$. 
The mixing matrix
has a form 
\begin{equation}
C^{(1)}=\left(\begin{array}{cccc}
X_1 & X_2 & 0 & 0\\
X_3 & X_4 & 0 & 0\\
0 & 0 & X_5 & X_6\\
0 & 0 & X_7 & X_8\end{array}\right)\:.
\end{equation}
The divergent contribution comes from
\begin{equation}
\tilde{Q}_{q,\wbar q}(z,\wbar{z})=X_8\: z\: v_{4}(1-z)
\wbar{v}_{4}(1-\wbar{z})+\ldots=X_8'\: 
z(1-z)^{-5/2}(1-\wbar{z})^{3/2}+\ldots\:,
\label{eq:Qsin}
\end{equation}
where in the second term of (\ref{eq:Qsin})
the exponents are shown for the $n_h=2$ case. 

Performing calculations similar to 
the case of odd $n_h$ we obtain for $n_h=2$
logarithmic singularity
\begin{equation}
\langle q,\wbar q|q,\wbar q\rangle\sim\lim_{\rho\to0}R^{6}
\frac{\log(\rho/R)}{\rho^{6}}+\ldots\,.
\end{equation}
For point-like solutions with higher
$n_{h} \in 2 \mathbb{Z} $ we obtain even stronger singularities,
\ie $\langle q,\wbar q|q,\wbar q\rangle\sim\lim_{\rho\to0}1/\rho^{6n_{h}-6}$.

To sum up, all point-like solutions constructed in Section
\ref{sec:psol}
are non-normalizable
because they do not have continuation in $\nu_h \in \mathbb{R}$
and their norm is singular at $\nu_h=0$.
Thus, they do not belong to the Hamiltonian spectrum 
and therefore they do not contribute to the physical scattering processes.

\subsection{Point-like solutions with infinite energy}
Since in the analytical continuation of the energy 
there are poles for odd $2 i \nu_h$ 
one can expect that 
according to (\ref{eq:hhsym})
this pole-solutions
have twin-solutions for $\nu_h \in \mathbb{R}$ and odd $n_h$.
Indeed, such point-like solutions also appear.

Let us consider the first pole with $q_3=q_4=0$, $n_h=0$ and $i \nu_h=3/2$, 
\ie $(h,\wbar h)=(2,2)$. It should have a
twin-solution with $(h,\wbar h)=(2,-1)$ and $q_3=q_4=0$.
Solving the differential equations (\ref{eq:Eq-1}) 
and resuming the series we obtain
\begin{equation}
\tilde Q_{q,\wbar q}(z,\wbar z)= \alpha_1 \frac{z(z+1)}{(1-z)^3}+
\alpha_2 \left(
\frac{z(z+1)}{(1-z)^3} \Log[z \wbar z]+
\frac{2z}{(1-z)^2}
\right) 
\label{eq:pol1}
\end{equation}
where $\alpha_3=\alpha_4=0$ in Eq.\ (\ref{eq:C0}) and $\alpha_1$, $\alpha_2$ 
are arbitrary constants. 
Using Eq.~(\ref{eq:Qz-symmetry}) one can find that solution proportional
to $\alpha_1$ have $\theta_4=\pi$ while
one at $\alpha_2$ have $\theta_4=0$.
Considering (\ref{eq:E-fin}) one cannot evaluate their energy,
\ie $E_4=\pm \infty$.

Similar solutions exist for other energy poles.
Applying Eq.~(\ref{eq:normzz}) to these pole-like solutions 
one gets the vanishing norm. 
Thus, these solutions also do not contribute to the physical processes. 

\section*{Summary}
\label{sec:sum}
We have focused on four-Reggeon exchanges with 
the non-vanishing conformal Lorentz spin, $n_h$.
These exchanges contribute to the elastic 
amplitude processes.
In this work the energy spectrum of the exchanged states as well as 
the spectrum of conformal charges have been calculated,
which enabled to calculate high energy behaviour of the scattering amplitude.
The Reggeized gluon states contribute also to
processes of deep inelastic scattering of a virtual  
photon $\gamma^\ast(Q^2)$ off
a hadron with mass $M^2$ where $\Lambda_{QCD} \ll M^2 \ll Q^2$. 
With the help of techniques presented in this paper it is possible
to calculate 
the anomalous dimensions of QCD for the structure function. 
This can be achieved by performing analytical continuation
of the energy $E_4(\nu_h)$ in to the complex $\nu_h-$plane and compute
expansion coefficients of the energy around its poles 
\cite{Jaroszewicz:1982gr,Korchemsky:2003rc}. Positions of these poles 
describe possible values of the twist 
while the coefficients set a dependence of
anomalous dimensions on the strong coupling constant.
To this end, the first step is the same as in the case of Eq.~(\ref{A}), 
\ie to solve the Schr\"odinger equation (\ref{eq:Schr})
using Baxter $Q-$operator method.

To find the energy spectrum (\ie intercepts) and
conformal charges, the Baxter eigenproblem was numerically solved.
It turns out that apart from the ordinary states, 
which 
form one dimensional trajectories in the space of
conformal charges along
$\nu_h$ direction, 
there exist other 
solutions to the Baxter equation, \ie the point-like solutions.
In order to contribute to the scattering amplitude processes, 
\ie to be physical, the pertinent 
solutions should be normalizable according to the 
$\SL(2,\mathbb{C})$ scalar product. 
Thus, the scalar product of 
the trajectory solutions with continuous parameters $\nu_h$ and $\nu_h'$
has to be proportional to $\delta(\nu_h-\nu_h') \delta_{n_h n_h'}$ 
while the scalar product 
of the point-like solutions to the Kronecker delta $\delta_{n_h n_h'}$.
On the other hand
by construction, the wave-functions of the eigenstates should have
transformation property w.r.t.~$\SL(2,\mathbb{C})$ conformal transformations, 
\ie they have to
be homogeneous functions of two-dimensional coordinates and should
not involve any scale. 
These conditions are not  met by discrete spectrum.
Since the discrete point-like solutions have infinite or vanishing norm,
they are non-normalizable and they do not belong to the 
physical spectrum
of the Hamiltonian. Moreover, they do not contribute to the physical 
processes. 

The point-like states have been considered in
Ref.~\cite{deVega:2002im,deVega:2002im-bis}
as candidates for the ground-states  which would give 
the main contribution to the scattering processes.
There are no point-like solutions
for the vanishing conformal Lorentz spin. 
However, such solutions appear for $n_h\ne 0$.

These discrete solutions appear due to the symmetry of Casimir operator
namely  $\wbar h \leftrightarrow 1- \wbar  h$ which may throw
solutions outside the physical region. This symmetry is valid only for 
(half-)integer conformal  weights $(h, \wbar h)$.
The main issue is how a given solution to the conditions 
(\ref{eq:symcon})
looks like in the vicinity of $h, \wbar h \in \mathbb{Z}/2$. 
If one solves the quantization conditions for $h=(1-n_h)/2$
and $\wbar h=(1+n_h)/2$ then one should be prepared to encounter both
physical, \ie normalizable solutions as well as non-physical 
non-normalizable ones. The only way to distinguish them is
to move a little bit away from these values of $h$ and $\wbar h$ and 
check if 
they are situated on some physical trajectory with $\nu_h \in \mathbb{R}$.

The only physical spectrum of the Hamiltonian (\ref{eq:sepH})
is continuous spectrum.
This spectrum may be parameterized using WKB approximation 
performed for large values of conformal charges.
The energy states depend on continuous parameter $\nu_h$, integer conformal
Lorentz spin $n_h$  and a set of integer parameters which enumerate the trajectories as a vertices of WKB lattices.
The energy of the states with $n_h\ne0$
is always higher than energy  of ground state in the $n_h=0$
sector. However $n_h \ne 0$ sectors may give the leading contribution to the scattering amplitude if the ground state does not couple to scattered particles.

We hope that this work explains the situation of four 
reggeized gluon solutions for $n_h \ne 0$.
Although the point-like solutions do not contribute to the scattering 
amplitudes
they are a very interesting part of the Baxter eigenproblem.
We have to remember about them solving $\SL(2,\mathbb{C})$ 
system in the future.
One can expect that the point-like solutions 
may also appear for $N\ne4$ Reggeon solutions.
Finally, we have to agree with \cite{Derkachov:2002wz} that 
in the multi-colour limit the four-Reggeon state with the lowest energy
which governs the four Reggeon physical scattering processes 
lies on the trajectory (\ref{eq:groundstate}) from the $n_h=0$ sector
with $q_3=0$ and $E_4=-2.69664$.

\section*{Acknowledgements}
I would like to warmly thank to
G.~P.~Korchemsky,
A.~N.~Manashov and
M.~Prasza{\l}owicz
for fruitful discussions and help during preparation of this work.
This work was supported by
the grant of the Polish Ministry of Science and Education 
P03B 024 27 (2004-2007).

\newpage

\appendix

\section{Recurrence relations for  $u_n^{(k)}$}
\subsection{Coefficients $u_n^{(k)}$ around $z=0$}
\label{ap:uk}

The coefficient for the recurrence relations around $z=0$
are defined as follows
\begin{equation}
a_{0,n}= -2 (n - s + 1)^4 -\sum_{k=2}^{4} i^k q_k (n-s+1)^{4-k}\,,
\label{eq:agA0}
\end{equation}
\begin{equation}
a_{-1,n}= (n-2s+1)^4\,,
\label{eq:agAm1}
\end{equation}
\begin{equation}
a_{1,n}= (n+1)^4\,,
\label{eq:agA1}
\end{equation}
while
\begin{equation}
a^{(k)}_{i,n}=\frac{d^k}{dn^k} a_{i,n}\,.
\label{eq:agAmk}
\end{equation}

Now, in Eq.~(\ref{eq:power-series-0}) $u_{-1}^{(k)}=0$,  $u_{0}^{(k)}=1$ and
$u_{m}^{(k)}\equiv u_{m}^{(k)}(q)$ is defined by
\begin{multline}
u_{m}^{(k)} a_{1,m-1}
=
-u_{m-1}^{(k)} a_{0,m-1}
-u_{m-2}^{(k)} a_{-1,m-1}\\
-\sum_{i=1}^{k-1}
\frac{1}{(k-i)!}
(u_{m}^{(i)} a_{1,m-1}^{(k-i)}
+u_{m-1}^{(i)} a_{0,m-1}^{(k-i)}
+u_{m-2}^{(i)} a_{-1,m-1}^{(k-i)})\,,
\label{eq:ff}
\end{multline}
with
\begin{equation}
u_k(z)=(k-1)! z^{1-s} \sum_{n=0}^{\infty} 
u_n^{(k)} z^n\,.
\label{eq:uk}
\end{equation}
Finally, one obtains Eq.~(\ref{eq:Q-0-h}).

\subsection{Coefficients $v_m^{(j)}$ around $z=1$}
\label{ap:vk}

The coefficient for the recurrence relations around $z=1$
can be defined as 
\begin{equation}
A_{1,n}=(-1 + n) n (2 - h + n - 4 s) (1 + h + n - 4 s)\,,
\label{eq:A41}
\end{equation}
\begin{multline}
A_{2,n}=-(n (6 + 3 n (5 + n (4 + n)) + i q_3 + h (1 + 2 n - 2 s) - 28 s 
        -  8 n (7 + 3 n) s \\
+ 8 (6 + 7 n) s^2 
	- 32 s^3 + h^2 (-1 - 2 n + 2 s)))\,,
\label{eq:A42}
\end{multline}
\begin{multline}
A_{3,n}=14 + n (39 + n (40 + 3 n (6 + n)) + i q_3) + i q_3 - q_4 
+ h (1 + n - s)^2 -   h^2 (1 + n - s)^2 \\
- 60 s - (4 n (34 + n (25 + 6 n)) + i q_3) s + 
  32 (1 + n) (3 + 2 n) s^2 - 4 (17 + 16 n) s^3 + 18 s^4\,,
\label{eq:A43}
\end{multline}
\begin{equation}
A_{4,n}=-(2 + n - 2 s)^4\,,
\label{eq:A44}
\end{equation}
while
\begin{equation}
A^{(k)}_{i,n}=\frac{d^k}{dn^k} A_{i,n}\,.
\label{eq:agAm1k}
\end{equation}

For non-$\Log(1-z)$ solutions we have recurrence relations for 
$v_m^{(i)}\equiv v_m^{(i)}(q)$ from Eq.~(\ref{eq:v-series}):
\begin{equation}
\sum_{k=0}^{3} v_{m-k}^{(i)} A_{k+1,m-k+b_i}=0\,,
\label{eq:rrnlvm}
\end{equation}
where $b_i$ are corresponding roots of indicial equation and 
$A_{k,m}$ are defined above.
The solutions with one-$\Log(1-z)$ term have coefficients $v_m^{(i)}$
defined as
\begin{equation}
\sum_{k=0}^{3} v_{m-k}^{(i)} A_{k+1,m-k+b_i}+
\sum_{k=0}^{3} v_{m-k-d_{ij}}^{(j)} A^{(1)}_{k+1,m-k+b_i}=0\,,
\label{eq:rr1lvm}
\end{equation}
where $d_{ij}=b_i-b_j$ while the index $j$ is related with the solutions
at one-$\Log(1-z)$ term (see Eq.~\ref{eq:v1q3s1c}). 
Finally, the coefficients of two-$\Log(1-z)$ solutions, \ie $v_m^{(i)}$
have following constraints
\begin{multline}
\sum_{k=0}^{3} v_{m-k}^{(i)} A_{k+1,m-k+b_i}+
\sum_{k=0}^{3} v_{m-k-d_{ij}}^{(j)} A^{(1)}_{k+1,m-k+b_i}\\
+\sum_{k=0}^{3} v_{m-k-d_{il}}^{(l)} A^{(2)}_{k+1,m-k+b_i}=0\,,
\label{eq:rr2lvm}
\end{multline}
where $b_l<b_j<b_i$ and the index $j$ is related with the solutions
at one-$\Log(1-z)$ term while $l$ corresponds to the solution at two-$\Log(1-z)$ term.
Moreover, one has to fix one coefficient for each solution, 
\ie usually $v_0^{(i)}=1$, and one coefficient for the each $\Log(1-z)$ term.

If $A^{i}_{k,m}=0$ for some $i,k,m$ then recurrence relations may be
closed. In this situation to calculate $v^{(k)}_{m}$ one has to solve
a set of linear equations. Moreover, these linear equation also can give
additional constraints on conformal charges. This is mechanism of 
point-like solution appearing. Such solutions
have usually less $\Log(1-z)$ terms
than normal one.


\end{document}